\documentclass{elsart}
\usepackage{graphics}
\usepackage{graphicx}

\usepackage{amssymb}
\usepackage{epsfig}
\usepackage{amsmath}
\usepackage{bm}

\begin{document}

\begin{frontmatter}



\title{Quantum effects in optomechanical systems}


\author[innsb]{C. Genes}, \author[potsd]{A. Mari}, \author[unicam]{D. Vitali}, and \author[unicam]{P. Tombesi}
\address[innsb]{Institute for Theoretical Physics, University of Innsbruck, and Institute for
Quantum Optics and Quantum Information,
Austrian Academy of Sciences, Technikerstrasse 25, A-6020 Innsbruck, Austria}
\address[potsd]{Institute of Physics and Astronomy, University of Potsdam,
14476 Potsdam, Germany}
\address[unicam]{Dipartimento di Fisica, Universit\`a di Camerino, via Madonna delle Carceri, I-62032, Camerino (MC) Italy}

\tableofcontents

\begin{abstract}
The search for experimental demonstrations of the quantum behavior of macroscopic mechanical resonators is a fastly growing field of
investigation and recent results suggest that the generation of quantum states of resonators with a mass at the microgram scale is within reach.
In this chapter we give an overview of two important topics within this research field: cooling to the motional
ground state, and the generation of entanglement involving mechanical, optical and atomic degrees of freedom.
We focus on optomechanical systems where the resonator is coupled to one or more driven cavity modes by the radiation pressure
interaction. We show that robust stationary entanglement between the mechanical resonator and the output
fields of the cavity can be generated, and that this entanglement can be transferred to atomic ensembles placed within the cavity.
These results show that optomechanical devices are interesting candidates for the realization of quantum memories and interfaces
for continuous variable quantum communication networks.

\end{abstract}

\end{frontmatter}

\section{Introduction}
\label{sec:intro}

Mechanical resonators at the micro- and nano-meter scale are widely employed for a large variety of applications, more commonly
as sensors or actuators
in integrated electrical, optical, and opto-electronical systems \cite{blencowe,roukpt,optexpr,markus}.
Modifications of the resonator motion can be detected
with high sensitivity by looking at the radiation (or electric current) which interacted with the resonator. For example, small masses
can be detected by measuring the frequency shift induced on the resonator, while tiny displacements (or weak forces inducing
such displacements) can be measured by detecting the corresponding phase shift of the light interacting with it \cite{roukpt}.
The resonators are always subject to thermal noise, which is due to the coupling with internal and/or external degrees
of freedom and is one of the main factors limiting the sensitivity of these devices. However, due to the progress in
nanofabrication techniques, the mechanical quality factor $\mathcal{Q}_m$
(which quantifies this undesired coupling to environmental degrees of freedom) is steadily improving, suggesting that in the near future
these devices will reach the regime in which their sensitivity is limited by the ultimate quantum limits set by the Heisenberg principle.
The importance of the limits imposed by quantum mechanics on the resonator motion was first pointed out by Braginsky and
coworkers \cite{bragbook} in the completely different context of \emph{massive} resonators
employed in the detection of gravitational waves \cite{ligo}. However, in recent years the quest for the experimental demonstration of genuine
quantum states of macroscopic mechanical resonators has spread well beyond the gravitational wave physics community and has attracted a wide interest.
In fact, the detection of an unambiguous signature of the quantum behavior of a macroscopic oscillator, with a mass at least of the order of a microgram,
would shed further light onto the quantum-classical boundary problem~\cite{found}. In fact, nothing in the principles of quantum mechanics
prevents macroscopic systems to be prepared in genuine quantum states. However, it is not yet clear how far one can go in this direction~\cite{Leggett},
and a complete understanding of how classical behavior emerges from the quantum substrate requires the design and the implementation of
dedicated experiments. Examples of this kind are single-particle interference of macro-molecules~\cite{Hackermueller},
the demonstration of entanglement between collective spins of atomic ensembles~\cite{Polzik},
and of entanglement in Josephson-junction qubits~\cite{Berkley}.
For what concerns mechanical resonators, the experimental efforts are currently focusing on cooling them down to their motional ground state \cite{roukpt}.
This goal has not been achieved yet, but promising results in this direction have been obtained in different setups
\cite{cohadon99,schwab,karrai,naik,arcizet06,gigan06,arcizet06b,bouwm,vahalacool,mavalvala,rugar,wineland,markusepl,sidebcooling,harris,lehnert,vinante,lehnert08,tobias09,markus09},
involving different examples of mechanical resonators coupled either to radiative or to electrical degrees of freedom.
Ground state cooling of microgram-scale resonators seems to be within reach, as already suggested by various theoretical proposals
\cite{Mancini98,brag,courty,vitalirapcomm,quiescence02,imazoller,tianzoller,marquardt,wilson-rae,genes07,dantan07,wilson-rae08} which showed how
a mechanical oscillator can be coupled to another system so that the latter can act as an effective zero-temperature reservoir.
In the first part of this chapter we shall review the problem of ground state cooling of a mechanical resonator, by focusing onto the case where
the role of effective zero-temperature ``fridge'' is played by an optical cavity mode,
coupled to the resonator by radiation pressure. In this case this interaction
can be exploited for cooling in two different ways: i) back-action, or self-cooling \cite{brag,marquardt,wilson-rae,genes07,dantan07,wilson-rae08}
in which the off-resonant operation
of the cavity results in a retarded back action on the mechanical
system and hence in a ``self''-modification of its dynamics
\cite{karrai,gigan06,arcizet06b,vahalacool,mavalvala,wineland,markusepl,sidebcooling,harris,lehnert,lehnert08,tobias09,markus09}; ii) cold-damping quantum
feedback, where the oscillator position is measured through a phase-sensitive detection of the cavity
output and the resulting photocurrent is used for a real-time correction of the dynamics \cite{cohadon99,arcizet06,bouwm,rugar,vinante}.
We shall compare the
two approaches and see that while back-action cooling is optimized in the good cavity limit where the resonator frequency is larger than the cavity bandwidth,
cold damping is preferable in the opposite regime of larger cavity bandwidths~\cite{genes07}. It should be noticed that the
model Hamiltonian based on radiation pressure coupling between an optical cavity mode and one movable cavity mirror is quite general and
immediately extendible to other situations, such as the toroidal microcavities of Refs.~\cite{vahalacool,sidebcooling}, the capacitively coupled
systems of Refs.~\cite{wineland,lehnert} and even atomic condensate systems \cite{gupta}.

From the theory side, the generation of other examples of quantum states of a micro-mechanical resonator has been also considered. The most relevant
examples are given by squeezed and resonator-field (or atoms) entangled states.
Squeezed states of nano-mechanical resonators \cite{blencowe-wyb} are potentially useful
for surpassing the standard quantum limit for position and force detection \cite{bragbook}, and could be generated in different ways, either by
coupling to a qubit \cite{squee1}, or by measurement and feedback schemes \cite{quiescence02,squee2}. Entanglement is instead the
characteristic element of quantum theory, because it is responsible for correlations between observables that cannot be understood on the basis
of local realistic theories \cite{Bell64}. For this reason, there has been an increasing interest in establishing the conditions under which
entanglement between macroscopic objects can arise. Relevant experimental demonstration in this directions are given by the entanglement between
collective spins of atomic ensembles~\cite{Polzik}, and between Josephson-junction qubits~\cite{Berkley}. Then, starting from the proposal of
Ref.~\cite{PRL02} in which two mirrors of a ring cavity are entangled by the radiation pressure of the cavity mode, many proposals involved
nano- and micro-mechanical resonators, eventually entangled with other systems. One could entangle a nanomechanical oscillator with a
Cooper-pair box \cite{Armour03}, while Ref.~\cite{eisert} studied how to entangle an array of nanomechanical oscillators. Further proposals
suggested to entangle two charge qubits \cite{zou1} or two Josephson junctions \cite{cleland1} via nanomechanical resonators, or to entangle two
nanomechanical resonators via trapped ions \cite{tian1}, Cooper pair boxes \cite{tian2}, or dc-SQUIDS \cite{nori}. More recently, schemes for
entangling a superconducting coplanar waveguide field with a nanomechanical resonator, either via a Cooper pair box within the waveguide
\cite{ringsmuth}, or via direct capacitive coupling \cite{Vitali07}, have been proposed.
After Ref.~\cite{PRL02}, other optomechanical systems have been proposed for entangling optical and/or mechanical modes by means of the
radiation pressure interaction. Ref.~\cite{Peng03} considered two mirrors of two different cavities illuminated with entangled light beams,
while Refs.~\cite{pinard-epl,paternostro,meystre,wipf} considered different examples of double-cavity systems in which entanglement either
between different mechanical modes, or between a cavity mode and a vibrational mode of a cavity mirror have been studied.
Refs.~\cite{prl07,jopa} considered the simplest scheme capable of generating stationary optomechanical entanglement, i.e., a single Fabry-Perot
cavity either with one \cite{prl07}, or both \cite{jopa}, movable mirrors.

In the second part of the chapter we shall focus on the generation of stationary entanglement
by starting from the Fabry-Perot model of Ref.~\cite{prl07}, which is remarkable for its simplicity and robustness against temperature,
and extend its study in various directions. In fact, entangled optomechanical systems could be profitably used
for the realization of quantum communication networks, in which the mechanical modes play the role of local nodes where quantum information can
be stored and retrieved, and optical modes carry this information between the nodes. Refs.~\cite{prltelep,jmo,Pir06} proposed a scheme of this
kind, based on free-space light modes scattered by a single reflecting mirror, which could allow the implementation of continuous variable (CV)
quantum teleportation \cite{prltelep}, quantum telecloning \cite{jmo}, and entanglement swapping \cite{Pir06}. Therefore, any quantum
communication application involves \emph{traveling output} modes rather than intracavity ones, and it is important to study how the
optomechanical entanglement generated within the cavity is transferred to the output field. Furthermore, by considering the output field, one
can adopt a multiplexing approach because, by means of spectral filters, one can always select many different traveling output modes originating
from a single intracavity mode. One can therefore manipulate a multipartite system, eventually possessing multipartite
entanglement. We shall develop a general theory showing how the entanglement between the mechanical resonator and optical output modes can be
properly defined and calculated  \cite{output08}. We shall see that, together with its output field, the single Fabry-Perot cavity system
of Ref.~\cite{prl07} represents the ``cavity version''
of the free-space scheme of Refs.~\cite{prltelep,jmo}. In fact, as it happens in this latter scheme, all the relevant dynamics induced by
radiation pressure interaction is carried by the two output modes corresponding to the first Stokes and anti-Stokes sidebands of the driving
laser. In particular, the optomechanical entanglement with the intracavity mode is optimally transferred to the output Stokes sideband mode,
which is however robustly entangled also with the anti-Stokes output mode. We shall see that the present Fabry-Perot cavity system is preferable
with respect to the free space model of Refs.~\cite{prltelep,jmo}, because entanglement is achievable in a much more accessible experimental
parameter region. We shall then extend the analysis to the case of a doubly-driven cavity mode. We shall see that a peculiar parameter regime exists
where the optomechanical system, owing to the combined action of the two driven modes, is always stable and is characterized by
robust entanglement between the resonator and the cavity output fields.

In the last Section we shall investigate the possibility to couple and entangle in a robust way
optomechanical systems to atomic ensembles, in order to achieve a strongly-coupled
hybrid multipartite system \cite{fam,ian-ham}. We shall see that this is indeed possible, especially when the atomic ensemble is resonant
with the Stokes sideband induced by the resonator motion. Such hybrid systems might represent an important candidate for the realization of CV
quantum interfaces within CV quantum information networks.

\section{Cavity optomechanics via radiation pressure} \label{sec1}

The simplest cavity optomechanical system consists of a Fabry-Perot cavity with one heavy, fixed mirror through which a laser of frequency
$\omega _{l}$ drives a cavity mode, and another light end-mirror of mass $m$ (typically in the micro or nanogram range), free to oscillate at
some mechanical frequency $\omega _{m}$. Our treatment is however valid also for other cavity geometries in which one has an optical mode
coupled by radiation pressure to a mechanical degree of freedom. A notable example is provided by silica toroidal optical microcavities which
are coupled to radial vibrational modes of the supporting structure~\cite{vahalacool,vahala1}. Radiation pressure typically excites several
mechanical degrees of freedom of the system with different resonant frequencies. However, a single mechanical mode can be considered when a
bandpass filter in the detection scheme is used \cite{Pinard} and coupling between the different vibrational modes can be neglected. One has to
consider more than one mechanical mode only when two close mechanical resonances fall within the detection bandwidth (see Ref.~\cite{duemodi}
for the effect of a nearby mechanical mode on cooling and entanglement). The
Hamiltonian of the system describes two harmonic oscillators coupled via the radiation pressure interaction, and reads \cite{GIOV01}
\begin{equation}
H=\hbar \omega _{c}a^{\dagger }a+\frac{1}{2}\hbar \omega _{m}(p^{2}+q^{2})-\hbar G_{0}a^{\dagger }aq+i\hbar \mathcal{E}(a^{\dagger }e^{-i\omega
_{l}t}-ae^{i\omega _{l}t}).  \label{Ham}
\end{equation}%
The first term describes the energy of the cavity mode, with lowering operator $a$ ($[a,a^{\dag }]=1$), frequency $\omega _{c}$ (and therefore
detuned by $\Delta_0=\omega_c-\omega_l$ from the laser), and decay rate $\kappa $. The second term gives the energy of the mechanical mode,
described by dimensionless position and momentum operators $q$ and $p$ ($%
[q,p]=i$). The third term is the radiation-pressure coupling of rate $%
G_{0}=(\omega _{c}/L)\sqrt{\hbar /m\omega _{m}}$, where $m$ is the effective mass of the mechanical mode \cite{Pinard}, and $L$ is an effective
length that depends upon the cavity geometry: it coincides with the cavity length
in the Fabry-Perot case, and with the toroid radius in the case of Refs.~%
\cite{vahalacool,vahala1}. The last term describes the input driving by a laser with frequency $\omega _{l}$, where $\mathcal{E}$ is related to
the input laser power $\mathcal{P}$ by $|\mathcal{E}|=\sqrt{2\mathcal{P}\kappa /\hbar \omega _{l}}$. One can adopt the single cavity mode
description of Eq.~(\ref{Ham}) as long as one drives only one cavity mode and the mechanical frequency $\omega _{m}$ is much smaller than the
cavity free spectral range $FSR\sim c/2L$. In this case, in fact, scattering of photons from the driven mode into other cavity modes is
negligible \cite{law}.
\subsection{Langevin equations formalism}
The dynamics are also determined by the fluctuation-dissipation processes affecting both the optical and the mechanical mode. They can be taken
into account in a fully consistent way \cite{GIOV01} by considering the following set of nonlinear QLE (quantum Langevin equations), written in
a frame rotating at $\omega _{l}$
\begin{eqnarray}
\dot{q}& =& \omega _{m}p, \label{nonlinlang1}\\
\dot{p}& = & -\omega _{m}q-\gamma _{m}p+G_{0}a^{\dag }a+\xi , \label{nonlinlang2}\\
\dot{a}& = & -(\kappa +i\Delta _{0})a+iG_{0}aq+\mathcal{E}+\sqrt{2\kappa }%
a^{in}. \label{nonlinlang3}
\end{eqnarray}
The mechanical mode is affected by a viscous force with damping rate $\gamma _{m}$ and by a Brownian stochastic force with zero mean value $\xi
(t) $, possessing the correlation function \cite{GIOV01,Landau}
\begin{equation}
\left\langle \xi (t)\xi (t^{\prime })\right\rangle =\frac{\gamma _{m}}{%
\omega _{m}}\int \frac{d\omega }{2\pi }e^{-i\omega (t-t^{\prime })}\omega %
\left[ \coth \left( \frac{\hbar \omega }{2k_{B}T_{0}}\right) +1\right] , \label{browncorre}
\end{equation}%
where $k_{B}$ is the Boltzmann constant and $T_{0}$ is the temperature of the reservoir of the micromechanical oscillator. The correlation
function and the commutator of the Gaussian stochastic force $\xi (t)$ are not proportional to a Dirac delta and therefore $\xi (t)$ is a
non-Markovian stochastic process. This fact guarantees that the QLE of Eqs.~(\ref{nonlinlang1})-(\ref{nonlinlang3}) preserve the correct
commutation relations between operators during the time evolution \cite{GIOV01}. However, a Markovian description of the symmetrized
correlations of $\xi(t)$ is justified in two different limits, which are both met in typical experimental situations: i) not too low
temperatures $k_{B} T_0/\hbar\omega_{m} \gg 1$, which for typical values is satisfied even at cryogenic temperatures; ii) high mechanical
quality factor ${\cal Q}=\omega_m/\gamma_m \to \infty$ \cite{benguria}, which is an important condition for the observation of quantum effects
on the mechanical resonator. In this case the correlation function of Eq.~(\ref{browncorre}) can be approximated as
\begin{equation}
\label{browncorre2}\left\langle \xi(t)\xi(t^{\prime})\right\rangle \simeq \gamma_{m}\left[  (2n_{0}+1) \delta(t-t^{\prime})+i
\frac{\delta^{\prime}(t-t^{\prime})}{\omega_{m}}\right]  ,
\end{equation}
where $n_{0}=\left( \exp \{\hbar \omega _{m}/k_{B}T_{0}\}-1\right) ^{-1}$ is the mean thermal excitation number of the resonator
and $\delta^{\prime}(t-t^{\prime})$ denotes the derivative of the Dirac delta.

The cavity mode amplitude instead decays at the rate $\kappa $ and is affected by the vacuum radiation input noise $a^{in}(t)$, whose
correlation functions are given by \cite{gard}
\begin{eqnarray}
\langle a^{in}(t)a^{in,\dag }(t^{\prime })\rangle &=&\left[ N(\omega _{c})+1%
\right] \delta (t-t^{\prime }). \label{input1}\\
\langle a^{in,\dag }(t)a^{in}(t^{\prime })\rangle &=& N(\omega _{c})\delta (t-t^{\prime }), \label{input2}
\end{eqnarray}
where $N(\omega _{c})=\left( \exp \{\hbar \omega _{c}/k_{B}T_{0}\}-1\right) ^{-1}$ is the equilibrium mean thermal photon number. At optical
frequencies $\hbar \omega _{c}/k_{B}T_{0}\gg 1$ and therefore $N(\omega _{c})\simeq 0$, so that only the correlation function of
Eq.~(\ref{input1}) is relevant.

Equations~(\ref{nonlinlang1})-(\ref{nonlinlang3}) are not easy to analyze owing to the nonlinearity. However, one can proceed with a
linearization of operators around the steady state. The semiclassical steady state is characterized by an intracavity field amplitude $\alpha
_{s}$ ($|\alpha _{s}|\gg 1$), and a new equilibrium position for the oscillator, displaced by $q_{s}$. The parameters $\alpha _{s}$ and $q_{s}$
are the solutions of the nonlinear algebraic equations obtained by factorizing Eqs.~(\ref{nonlinlang1})-(\ref{nonlinlang3}) and setting the time
derivatives to zero:
\begin{eqnarray}
q_{s}&=&\frac{G_{0}|\alpha _{s}|^{2}}{\omega _{m}}, \\
\alpha _{s}&=&\frac{\mathcal{E}}{\kappa +i\Delta },
\end{eqnarray}
where the latter equation is in fact the nonlinear equation determining $\alpha _{s}$, since the effective cavity detuning $\Delta $, including
radiation pressure effects, is given by \cite{manc-tomb}
\begin{equation}
\Delta =\Delta _{0}-\frac{G_{0}^{2}|\alpha _{s}|^{2}}{\omega _{m}}.
\end{equation}
Rewriting each Heisenberg operator of Eqs.~(\ref{nonlinlang1})-(\ref{nonlinlang3}) as the c-number steady state value plus an additional
fluctuation operator with zero mean value, one gets the exact QLE for the fluctuations
\begin{eqnarray}
\delta \dot{q}& =&\omega _{m}\delta p, \\
\delta \dot{p}& =&-\omega _{m}\delta q-\gamma _{m}\delta p+G_{0}\left( \alpha _{s}\delta a^{\dag }+\alpha _{s}^{\ast }\delta a\right) +\delta
a^{\dag
}\delta a+\xi , \\
\delta \dot{a}& =&-(\kappa +i\Delta )\delta a+iG_{0}\left( \alpha _{s}+\delta a\right) \delta q+\sqrt{2\kappa }a^{in}. \label{mode}
\end{eqnarray}
Since we have assumed $|\alpha _{s}|\gg 1$, one can safely neglect the nonlinear terms $\delta a^{\dag }\delta a$ and $\delta a\delta q$ in the
equations above, and get the linearized QLE
\begin{eqnarray}
\delta \dot{q}& =&\omega _{m}\delta p, \label{lle1}\\
\delta \dot{p}& =&-\omega _{m}\delta q-\gamma _{m}\delta p+G\delta X+\xi , \label{lle2}\\
\delta \dot{X}& =&-\kappa \delta X+\Delta \delta Y+\sqrt{2\kappa }X^{in}, \label{lle3}\\
\delta \dot{Y}& =&-\kappa \delta Y-\Delta \delta X+G\delta q+\sqrt{2\kappa } Y^{in}. \label{lle4}
\end{eqnarray}
Here we have chosen the phase reference of the cavity field so that $\alpha _{s}$ is real and positive, we have defined the cavity field
quadratures $\delta X\equiv (\delta a+\delta a^{\dag })/\sqrt{2}$ and $\delta Y\equiv (\delta a-\delta a^{\dag })/i\sqrt{2}$ and the
corresponding Hermitian input
noise operators $X^{in}\equiv (a^{in}+a^{in,\dag })/\sqrt{2}$ and $%
Y^{in}\equiv (a^{in}-a^{in,\dag })/i\sqrt{2}$. The linearized QLE show that the mechanical mode is coupled to the cavity mode quadrature
fluctuations by the effective optomechanical coupling
\begin{equation}
G=G_{0}\alpha _{s}\sqrt{2}=\frac{2\omega _{c}}{L}\sqrt{\frac{\mathcal{P}%
\kappa }{m\omega _{m}\omega _{l}\left( \kappa ^{2}+\Delta ^{2}\right) }}, \label{optoc}
\end{equation}%
which can be made very large by increasing the intracavity amplitude $\alpha _{s}$. Notice that together with the condition $\omega _{m}\ll c/L$
which is required for the single cavity mode description, $|\alpha _{s}|\gg 1$ is the \emph{only} assumption required by the linearized
approach. This is in
contrast with the perturbative approaches described in \cite{wilson-rae}, where a reduced master equation of the mechanical resonator is
derived under the weak-coupling assumption $G\ll \omega
_{m}$.

\subsection{Stability analysis}

\begin{figure}[tb]
\centerline{\includegraphics[width=0.95\textwidth]{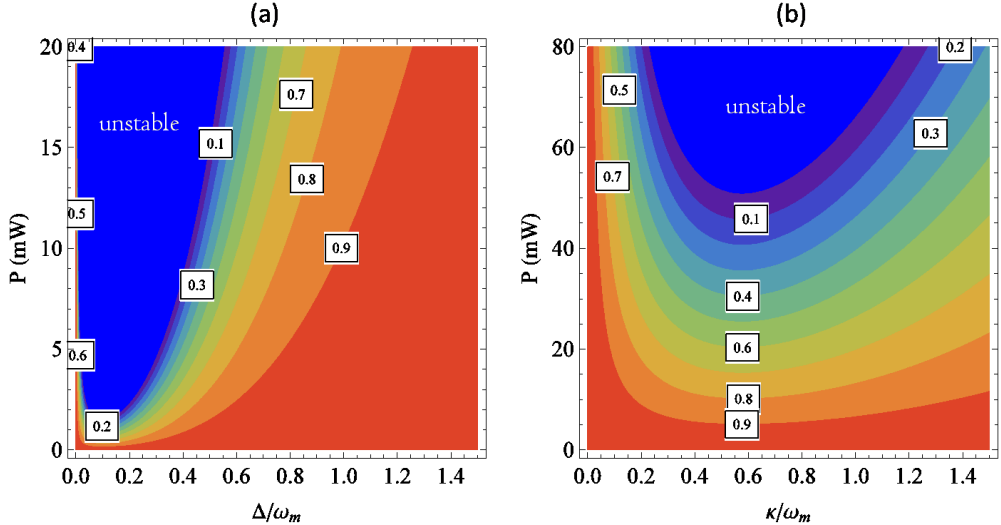}} \caption{Stability condition in the
red-detuning region. (a) Contour plot of the stability parameter $\protect\eta $ of Eq.~(\protect\ref{eta})
as a function of input power $\mathcal{P}$ and normalized detuning $%
\Delta /\protect\omega _{m}$. The parameter set $p_{0}$=($\omega_{m},Q_{m},m,L,\lambda_{c},T_{0})=(2\pi \times 10$ MHz, $10^{5}$, $30$ ng, $0.5$
mm, $1064$ nm, $0.6$ K has been used, together with $\mathcal{F}=8\times 10^{4}$ (corresponding to $\protect\kappa %
=0.37 \protect\omega _{m}$). The blue area corresponds to the unstable regime. (b) Stability parameter $\protect\eta $ versus
$\mathcal{P}$
and the normalized cavity decay rate $\protect\kappa /\protect\omega _{m}$ at $%
\Delta =\protect\omega _{m}$.}\label{stability}
\end{figure}

The stability analysis can be performed on the linearized set of equations Eqs.~(\ref{nonlinlang1})-(\ref{nonlinlang3}) by using the
Routh-Hurwitz criterion \cite{grad}. Two conditions are obtained
\begin{eqnarray}
s_{1}&=&2\gamma _{m}\kappa \left\{ \left[ \kappa ^{2}+\left( \omega _{m}-\Delta \right) ^{2}\right] \left[ \kappa ^{2}+\left( \omega _{m}+\Delta
\right) ^{2}\right] \right.  \label{stab1}\\
& + & \left. \gamma _{m}\left[ \left( \gamma _{m}+2\kappa \right) \left( \kappa ^{2}+\Delta ^{2}\right) +2\kappa \omega _{m}^{2}\right] \right\}
+\Delta
\omega _{m}G^{2}\left( \gamma _{m}+2\kappa \right) ^{2}>0, \\
 s_{2}&=&\omega _{m}\left( \kappa ^{2}+\Delta ^{2}\right) -G^{2}\Delta >0. \label{stab2}
\end{eqnarray}
The violation of the first condition, $s_{1}< 0$, indicates instability in the domain of blue-detuned laser ($\Delta <0$) and it corresponds to
the emergence of a self-sustained oscillation regime where the mirror effective damping rate vanishes. In this regime, the laser field energy
leaks into field harmonics at frequencies $\omega _{l}\pm r\omega _{m}\,$($r=1,2...$) and also feeds the mirror coherent oscillations. A complex
multistable regime can emerge as described in \cite{harris06}. The violation of the second condition $s_{2}< 0$ indicates the emergence of the
well-known effect of bistable behavior observed in \cite{dorsel} and occurs only for positive detunings ($\Delta >0$). In the following we
restrict our analysis to positive detunings in the stable regime where both $s_{1}$ and $s_{2}$ conditions are fulfilled. A parametric plot
showing the domain of stability in the red-detuning regime $\Delta >0$ is shown in Fig.~\ref{stability} where we have plotted the stability
parameter
\begin{equation}
\eta =1-\frac{G^{2}\Delta }{\omega _{m}\left( \kappa ^{2}+\Delta ^{2}\right) }.  \label{eta}
\end{equation}%
Negative values of $\eta $ indicate the emergence of instability. We have chosen the following set of parameters which will be used extensively
throughout the chapter and which is denoted by $p_{0}$=($\omega_{m},Q_{m},m,L,\lambda_{c},T_{0})=(2\pi \times 10$ MHz, $10^{5}$, $30$ ng, $0.5$
mm, $1064$ nm, $0.6$ K). These values are comparable to those used in recent experiments
\cite{gigan06,arcizet06b,bouwm,markusepl,sidebcooling,harris,tobias09,markus09}.

\subsection{Covariance matrix and logarithmic negativity}

The mechanical and intracavity optical mode form a bipartite continuous variable (CV) system. We are interested in the properties of its steady
state which, due to the linearized treatment and to the Gaussian nature of the noise operators, is a zero-mean Gaussian state, completely
characterized by its symmetrized covariance matrix (CM). The latter is given by the $
4\times 4$ matrix with elements%
\begin{equation}
\mathcal{V}_{lm}=\frac{\left\langle u_{l}\left( \infty \right) u_{m}\left( \infty \right) +u_{m}\left( \infty \right) u_{l}\left( \infty \right)
\right\rangle }{2},  \label{CM}
\end{equation}%
where $u_{m}(\infty)$ is the asymptotic value of the $m$-th component of the vector of quadrature fluctuations
\begin{equation}
u(t)=\left( \delta q(t),\delta p(t),\delta X(t),\delta Y(t)\right) ^{\intercal }.
\end{equation}%
Its time evolution is given by Eqs.~(\ref{lle1})-(\ref{lle4}), which can be rewritten in compact form as
\begin{equation}
\frac{d}{dt}u(t)=Au(t)+v(t),
\end{equation}%
with $A$ the drift matrix
\begin{equation}
A=\left(
\begin{array}{cccc}
0 & \omega _{m} & 0 & 0 \\
-\omega _{m} & -\gamma _{m} & G & 0 \\
0 & 0 & -\kappa  & \Delta  \\
G & 0 & -\Delta  & -\kappa
\end{array}%
\right) ,
\end{equation}%
and $v(t)$ the vector of noises
\begin{equation}
v(t)=\left( 0,\xi (t),\sqrt{2\kappa } X^{in}(t),\sqrt{2\kappa } Y^{in}(t)\right) ^{\intercal }.
\end{equation}
The steady state CM can be determined by solving the Lyapunov equation
\begin{equation}
A\mathcal{V}+\mathcal{V}A^{\intercal }=-D,  \label{Lyapunov}
\end{equation}%
where $D$ is the $4\times 4$ diffusion matrix which characterizes the noise correlations and is defined by the relation $\left\langle
n_{l}\left(t \right) n_{m}\left( t^{\prime} \right) +n_{m}\left( t^{\prime} \right) n_{l}\left(t \right) \right\rangle /2= D_{lm}\delta(t-
t^{\prime})$. Using Eqs.~(\ref{browncorre2})-(\ref{input1}), $D$ can be written as
\begin{equation}
D=\mathrm{diag}[0,\gamma _{m}\left( 2n_{0}+1\right) ,\kappa ,\kappa ].
\end{equation}%
Eq.~(\ref{Lyapunov}) is a linear equation for $\mathcal{V}$ and it can be straightforwardly solved, but the general exact expression is very
cumbersome and will not be reported here.

The CM allows to calculate also the entanglement of the steady state. We adopt as entanglement measure the logarithmic negativity
$E_{\mathcal{N}}$, which is defined as \cite{logneg}
\begin{equation}\label{eq:logneg}
E_{\mathcal{N}}=\max [0,-\ln 2\eta ^{-}].
\end{equation}%
Here $\eta ^{-}\equiv 2^{-1/2}\left[ \Sigma (\mathcal{V})-\left[ \Sigma (%
\mathcal{V})^{2}-4\det \mathcal{V}\right] ^{1/2}\right] ^{1/2}$and $\Sigma (%
\mathcal{V})\equiv \det \mathcal{V}_{1}+\det \mathcal{V}_{2}-2\det \mathcal{V%
}_{c}$, with $\mathcal{V}_{1},\mathcal{V}_{2}$ and $\mathcal{V}_{c}$ being $2\times 2$ block matrices of
\begin{equation}
\mathcal{V}\equiv \left(
\begin{array}{cc}
\mathcal{V}_{1} & \mathcal{V}_{c} \\
\mathcal{V}_{c}^{T} & \mathcal{V}_{2}%
\end{array}%
\right) .  \label{CMatrix}
\end{equation}%
A bimodal Gaussian state is entangled if and only if $\eta ^{-}<1/2$, which is equivalent to Simon's necessary and sufficient entanglement
non-positive partial transpose criterion for Gaussian states \cite{simon}, which can be written as $4\det \mathcal{V}<\Sigma (\mathcal{V})-1/4$.
Logarithmic negativity is a convenient entanglement measure because it is the only one which can always be explicitly computed and it is also
additive. The drawback of $E_{\mathcal{N}}$ is that, differently from the entanglement of formation and the distillable entanglement, it is not
strongly super-additive and therefore it cannot be used to provide lower-bound estimates of the entanglement of a generic state by evaluating
the entanglement of Gaussian state with the same correlation matrix \cite{Wolf06}. This fact however is not important in our case because the
steady state of the system is Gaussian within the validity limit of our linearization procedure.

\section{Ground state cooling}

The steady state CM $\mathcal{V}$ determines also the mean energy of the mechanical resonator, which is given by
\begin{equation} \label{meanenergy}
U=\frac{\hbar \omega _{m}}{2}\left[ \left\langle \delta q^{2}\right\rangle +\left\langle \delta p^{2}\right\rangle \right] = \frac{\hbar \omega
_{m}}{2}\left[ \mathcal{V}_{11} +\mathcal{V}_{22} \right] \equiv \hbar \omega _{m}\left( n+\frac{1}{2}\right) ,
\end{equation}%
where $n=\left( \exp \{\hbar \omega _{m}/k_{B}T\}-1\right) ^{-1}$ is the occupancy corresponding to a bath temperature $T$. Obviously, in the
absence of coupling to the cavity field it is $n=n_{0}$, where $n_{0}$ corresponds to the actual temperature of the environment $T_{0}$. The
optomechanical coupling with the cavity mode can be used to 'engineer' an effective bath of much lower temperature $T\ll T_{0}$, so that the
mechanical resonator is cooled. Let us see when it is possible to reach the ideal condition $n\ll 1$, which corresponds to ground state cooling.

\subsection{Feedback cooling}

\begin{figure}[tb]
\centerline{\includegraphics[width=0.95\textwidth]{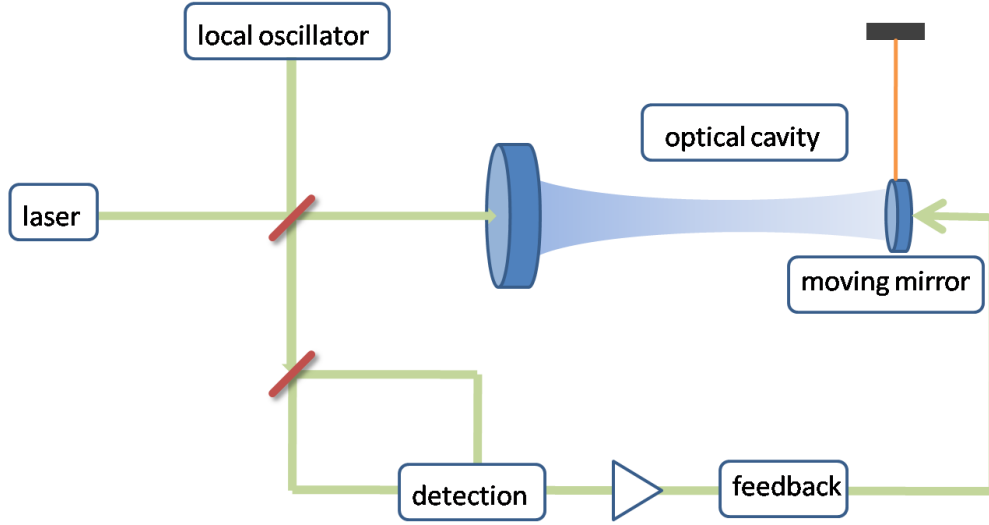}} \caption{Setup for feedback cooling (cold damping). The
cavity output field is homodyne detected (thus acquiring information about the mirror position) and a force proportional to its derivative
is fed back to the mirror.} \label{feedback_scheme}
\end{figure}

A simple way for cooling an object is to continuously detect its momentum and apply `corrective kicks' that continuously reduce it eventually to
zero \cite{Mancini98,vitalirapcomm,quiescence02}. This is the idea of feedback cooling illustrated in Fig.~\ref{feedback_scheme} where the mirror position is detected via phase-sensitive
homodyne detection of the cavity output field and a force proportional to the time derivative of the output signal (thus to the velocity) is fed
back to it.
By Fourier transforming Eq.~(\ref{lle4}) one obtains
\begin{equation} \label{fourierfeed}
\delta Y(\omega )=\frac{G(\kappa -i\omega )}{(\kappa -i\omega )^{2}+\Delta ^{2}}\delta q(\omega )+\mathrm{noise \; terms},
\end{equation}%
which shows that the intracavity phase-quadrature is sensitive to the mirror motion and moreover its optimal sensitivity is reached at
resonance, when $\Delta =0$. In this latter condition $\delta X(\omega )$ is not sensitive to the mirror motion, suggesting that the strongest
feedback effect is obtained by detecting the output phase-quadrature $Y^{out}$ and feeding it back to the resonator.

\subsubsection{Phase quadrature feedback}

As a consequence we set $\Delta =0$ and add a feedback force in Eq.~(\ref{lle2}) so that
\begin{equation}
\delta \dot{p}=-\omega _{m}\delta q-\gamma _{m}\delta p+G\delta X+\xi -\int_{-\infty }^{t}dsg(t-s)\delta Y^{est}(s),
\end{equation}%
where $Y^{est}(s)$ is the estimated intracavity phase-quadrature, which, using input-output relations \cite{gard} and focusing onto the ideal
scenario of perfect detection, is given by
\begin{equation}
\delta Y^{est}(t)=\frac{Y^{out}(t)}{\sqrt{2\kappa }}=\delta Y(t)-\frac{%
Y^{in}(t)}{\sqrt{2\kappa }}.
\end{equation}%
The filter function $g(t)$ is a causal kernel and $g(\omega)$ is its Fourier transform. We choose a simple standard derivative high-pass filter
\begin{equation}
g(t)=g_{cd}\frac{d}{dt}\left[ \theta (t)\omega _{fb}e^{-\omega _{fb}t}\right] \;\;\;\;\;\;\;g(\omega )=\frac{-i\omega g_{cd}}{1-i\omega /\omega
_{fb}}, \label{high-pass-filter}
\end{equation}%
so that $\omega _{fb}^{-1}$ plays the role of the time delay of the feedback loop, and $g_{cd}>0$ is the feedback gain. The ideal derivative
limit is obtained when $\omega _{fb}\rightarrow \infty $, implying $g(\omega )=-i\omega g_{cd}$ and therefore $g(t)=g_{cd}\delta ^{\prime }(t)$.
In this limit the feedback force is equal (apart from an additional noise term) to $-g_{cd}\delta \dot{Y}$ which, due to
Eq.~(\ref{fourierfeed}), is an additional viscous force $-(g_{cd}G/\kappa) \delta \dot{q}$ only in the bad cavity limit $\kappa \gg
\omega_m,\gamma_m$.

One can solve the Langevin equations supplemented with the feedback term in the Fourier domain. In fact, the two steady state oscillator
variances $\left\langle \delta q^{2}\right\rangle $ and $\left\langle \delta p^{2}\right\rangle $ can be expressed by the following frequency
integrals
\begin{equation}
\left\langle \delta q^{2}\right\rangle =\int_{-\infty }^{\infty }\frac{%
d\omega }{2\pi }S_{q}^{cd}(\omega ),\;\;\;\;\left\langle \delta
p^{2}\right\rangle =\int_{-\infty }^{\infty }\frac{d\omega }{2\pi }\frac{%
\omega ^{2}}{\omega _{m}^{2}}S_{q}^{cd}(\omega ),  \label{spectra}
\end{equation}%
where $S_{q}^{cd}(\omega )$ is the position noise spectrum. Its explicit expression is given by
\begin{equation}
S_{q}^{cd}(\omega )=|\chi _{eff}^{cd}(\omega )|^{2}[S_{th}(\omega )+S_{rp}(\omega)+S_{fb}(\omega )],
\end{equation}
where the thermal, radiation pressure and feedback-induced contributions are respectively given by
\begin{eqnarray}
S_{th}(\omega )& = &\frac{\gamma _{m}\omega }{\omega _{m}}\coth \left( \frac{%
\hbar \omega }{2k_{B}T_{0}}\right) , \\
S_{rp}(\omega )& =&\frac{G^{2}\kappa }{\kappa ^{2}+\omega ^{2}}, \\
S_{fb}(\omega )& =& \frac{|g(\omega )|^{2}}{4\kappa }
\end{eqnarray}
and $\chi _{eff}^{cd}(\omega )$ is the susceptibility of the mechanical oscillator modified by the feedback
\begin{equation}
\chi _{eff}^{cd}(\omega )=\omega _{m}\left[ \omega _{m}^{2}-\omega
^{2}-i\omega \gamma _{m}+\frac{g(\omega )G\omega _{m}}{\kappa -i\omega }%
\right] ^{-1}.  \label{chieffcd}
\end{equation}%
This effective susceptibility contains the relevant physics of cold damping. In fact it can be rewritten as the susceptibility of an harmonic
oscillator with effective (frequency-dependent) damping and oscillation frequency. The modification of resonance frequency (optical spring
effect \cite{mavalvala,quiescence02}) is typically small for the chosen parameter regime ($\omega_m \simeq 1$ MHz)
and the only relevant effect of feedback
is the modification of the mechanical damping which, in the case of the choice of Eq.~(\ref{high-pass-filter}), is given by
\begin{equation}
\gamma _{m}^{eff,cd}(\omega )=\gamma _{m}+\frac{g_{cd}G\omega _{m}\omega _{fb}(\kappa \omega _{fb}-\omega ^{2})}{(\kappa ^{2}+\omega
^{2})(\omega _{fb}^{2}+\omega ^{2})}.
\end{equation}%
This expression shows that the damping of the oscillator may be significantly increased due to the combined action of feedback and of radiation
pressure coupling to the field. In the ideal limit of instantaneous feedback and of a bad cavity, $\kappa,\omega_{fb} \gg \omega_m, \gamma_m$,
effective damping is frequency-independent and given by $\gamma _{m}^{eff,cd}\simeq \gamma _{m}+g_{cd}G\omega _{m}/\kappa = \gamma
_{m}(1+g_{2})$, where we have defined the scaled, dimensionless feedback gain $g_{2}\equiv g_{cd}G\omega _{m}/\kappa \gamma _{m}$
\cite{quiescence02}.

The presence of cold-damping feedback also modifies the stability conditions. The Routh-Hurwitz criteria are equivalent to the conditions that
all the poles of the effective susceptibility of Eq.~(\ref{chieffcd}) are in the lower complex half-plane. For the choice of
Eq.~(\ref{high-pass-filter}) there is only one non-trivial stability condition, which reads
\begin{eqnarray}
&&s_{cd}=\left[ \gamma _{m}\kappa \omega _{fb}+g_{cd}G\omega _{m}\omega
_{fb}+\omega _{m}^{2}(\kappa +\omega _{fb})\right]  \left[ (\kappa +\gamma _{m})(\kappa +\omega _{fb})(\gamma
_{m}+\omega _{fb})\right. \nonumber \\
&&\left. +\gamma _{m}\omega _{m}^{2}-g_{cd}G\omega _{m}\omega _{fb}%
\right]  -\kappa \omega _{m}^{2}\omega _{fb}(\kappa +\gamma _{m}+\omega _{fb})^{2}>0.  \label{stabcd}
\end{eqnarray}
This condition shows that the system may become unstable for large gain and finite feedback delay-time and cavity bandwidth because in this limit the
feedback force can be out-of-phase with the oscillator motion and become an accelerating rather than a viscous force \cite{genes07}.

\begin{figure}[tb]
\centerline{\includegraphics[width=0.95\textwidth]{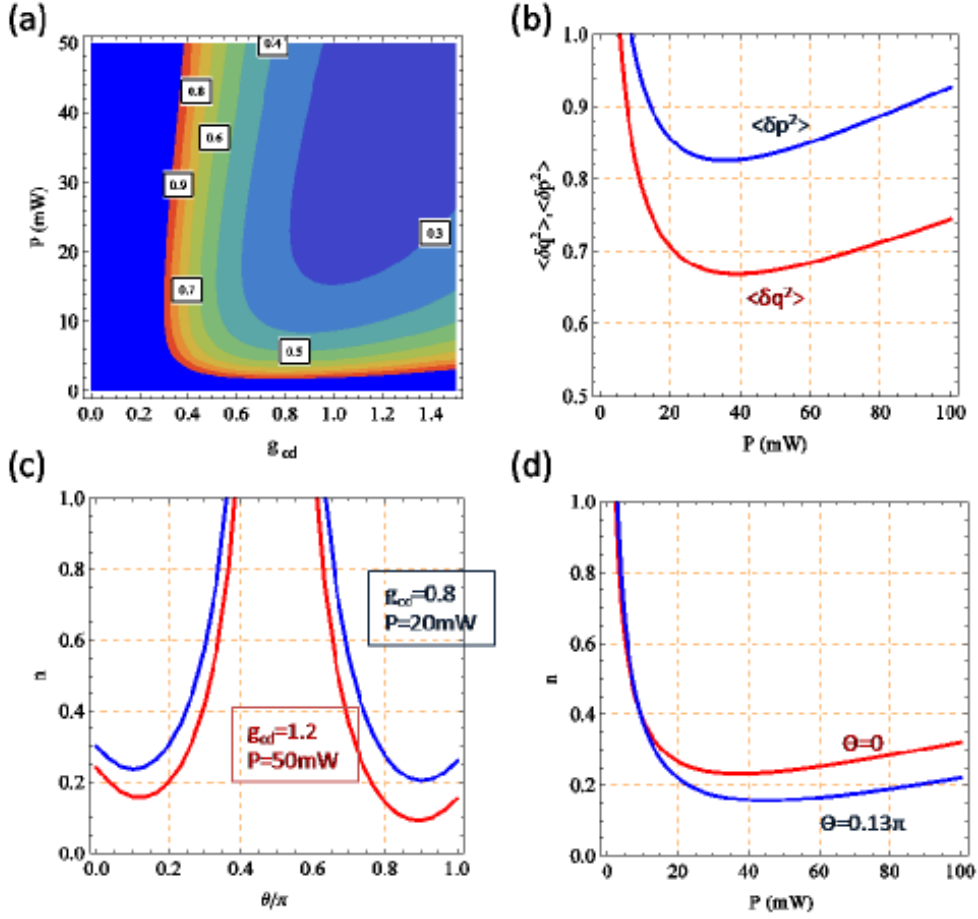}} \caption{Feedback cooling. (a) Contour plot
of $n$ as a function of $\mathcal{P}$ and $g_{cd}$. The parameters are $p_{0}$, $%
\protect\kappa =5 \protect\omega _{m}$ and $\protect\omega _{fb}=3.5%
\protect\omega _{m}$. (b) Illustration of the violation of energy equipartition around the optimal cooling regime. Parameters as before with $%
g_{cd}=1.2$. (c) $n$ versus the phase of the generalized quadrature $\theta$ for two sets of $g_{cd}$ and $%
\mathcal{P}$: the (upper) blue curve corresponds to $g_{cd}=0.8$ and $\mathcal{P}=20$ mW, while the (lower) red curve corresponds to
$g_{cd}=1.2$ and $\mathcal{P}=50$ mW. (d) Comparison of $n$ versus the input power $\mathcal{P}$ between the case of standard cold damping
feedback $\theta =0$ (upper red curve) and at a generalized detected quadrature with phase $\theta =0.13\pi$ (lower blue curve). Parameters as
before, with $g_{cd}=1.2$.}\label{feedback_cooling}
\end{figure}

The performance of cold-damping feedback for reaching ground state cooling is analyzed in detail in Ref.~\cite{genes07}, which shows that the
optimal parameter regime is $\kappa \gg \omega_{fb} \sim \omega_m \gg \gamma_m$, which correspond to a bad-cavity limit and a finite-bandwidth
feedback, i.e., with a feedback delay-time comparable to the resonator frequency. One gets in this case
\begin{eqnarray}
\left\langle \delta q^{2}\right\rangle  &\simeq & \left[1+g_{2}+\frac{\omega
_{fb}^{2}}{\omega _{m}^{2}}\right]^{-1}\left[\frac{g_{2}^{2}}{%
4\zeta }+\left( n_0+\frac{1}{2}+\frac{\zeta }{4}\right) \left( 1+%
\frac{\omega _{m}^{2}}{\omega _{fb}^{2}}\right) \right]  \label{qusqcd2} \\
\left\langle \delta p^{2}\right\rangle  &\simeq &\left[ 1+g_{2}+\frac{\omega
_{m}^{2}}{\omega _{fb}^{2}}\right] ^{-1}\left[ \frac{g_{2}^{2}}{4 \zeta }%
\left( 1+\frac{g_{2}\gamma _{m}\omega _{fb}}{\omega _{m}^{2}}\right) \right.
\notag \\
&+&\left. \left( n_0+\frac{1}{2}+\frac{\zeta }{4}\right) \left( 1+\frac{%
\omega _{m}^{2}}{\omega _{fb}^{2}}+\frac{g_{2}\gamma _{m}}{\omega _{fb}}%
\right) \right] ,  \label{pisqcd2}
\end{eqnarray}%
where we have defined the scaled dimensionless input power $%
\zeta=2G^{2}/\kappa\gamma_m$. These two expressions show that with cold-damping feedback, $\left\langle \delta q ^2\right\rangle \neq
\left\langle \delta p ^2\right\rangle$, i.e., energy equipartition does not hold anymore. The best cooling regime is achieved for $\omega _{fb}
\sim 3\omega _{m}$ and $g_2 \simeq \xi$ (i.e. $g_{cd}\simeq 2G/\omega _{m}$), i.e. for \emph{large but finite}
feedback gain~\cite{vitalirapcomm,quiescence02,genes07}. This is consistent
with the fact that stability imposes an upper bound to the feedback gain when $\kappa $ and $\omega _{fb}$ are finite. The optimal cooling
regime for cold damping is illustrated in Fig.~\ref{feedback_cooling}a, where $n$ is plotted versus the feedback gain $g_{cd}$ and the input
power $P$, at fixed $ \kappa =5\omega_m$ (bad-cavity condition) and $\omega _{fb}=3.5\omega_m$. Fig.~\ref{feedback_cooling}b instead explicitly
shows the violation of the equipartition condition even in this regime close to ground state (the feedback gain is fixed at the value
$g_{cd}=1.2$): the resonator is in a position-squeezed thermal state corresponding to a very low effective temperature.

\subsubsection{Generalized quadrature feedback}

The above analysis shows that cold-damping feedback better cools the mechanical resonator when the feedback is not instantaneous and therefore
the feedback force is not a simple viscous force. This suggests that one can further optimize feedback cooling by considering a
\emph{generalized} estimated quadrature which is a combination of phase and amplitude field quadratures. In fact one may expect that in the
optimal regime, the information provided by the amplitude quadrature $X^{out}(t)$ is also useful.

Therefore, in order to optimize cooling via feedback, we apply a feedback force involving a generalized estimated quadrature
\begin{equation}
\delta Y_{\theta }^{est}(t)=\frac{Y^{out}(t)\cos \theta +X^{out}(t)\sin \theta }{\sqrt{2\kappa }},
\end{equation}%
which is a linear combination of $Y^{out}(t)$ and $X^{out}(t)$ and where $\theta$ is a detection phase which has to be optimized. The adoption
of the new estimated quadrature leads to three effects: i) a modification of the expression for $\chi _{eff}^{cd}(\omega )$ of
Eq.~(\ref{chieffcd}) where $g(\omega )$ is replaced by $g(\omega )\cos \theta $; ii) a consequent reduction of the feedback-induced shot noise
term $S_{fb}(\omega )$; iii) a reduction of radiation pressure noise. In fact, the radiation pressure and feedback-induced noise contributions
become
\begin{eqnarray}
S_{rp}^{\theta }(\omega )& =&\frac{G^{2}\kappa }{\kappa ^{2}+\omega ^{2}} \left\vert 1-\frac{g(\omega )\sin \theta }{2G\kappa }\left( \kappa
+i\omega
\right) \right\vert ^{2}, \\
S_{fb}^{\theta }(\omega )& =& \frac{|g(\omega )|^{2}}{4\kappa }\cos ^{2}\theta .
\end{eqnarray}
An improvement over the standard cold-damping feedback scheme can be obtained when the shot noise reduction effect predominates over the
reduction of the effective damping due to feedback. This can be seen in Fig. (\ref{feedback_cooling}c) where for two different choices for
$g_{cd}$ and $\mathcal{P}$, the occupancy $n$ is plotted versus $\theta$. For one of these optimal phases, $\theta _{opt}=0.13\pi $, we plot in
Fig.~(\ref{feedback_cooling}d) $n$ as a function of $\mathcal{P}$ and compare it with the results of the standard phase quadrature feedback to
conclude that improvement via detection of a rotated output quadrature is indeed possible.

\subsection{Back-action cooling}

\begin{figure}[tb]
\centerline{\includegraphics[width=0.95\textwidth]{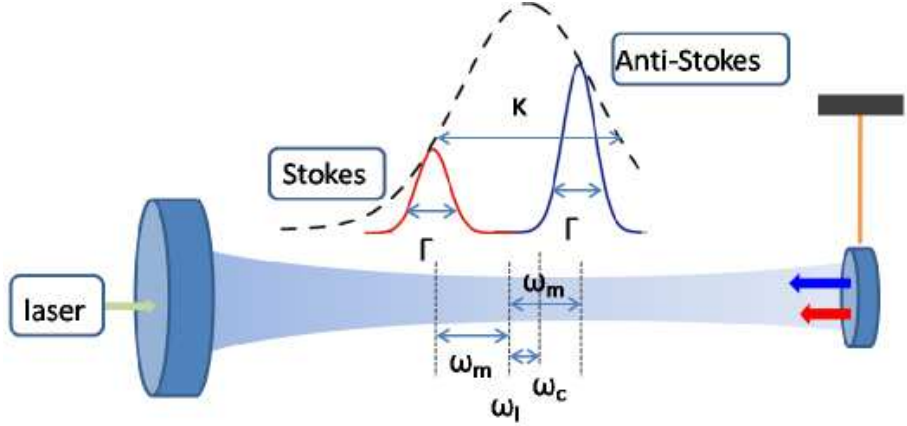}} \caption{Setup for cavity backaction cooling. Optical
sidebands are scattered unevenly by the moving mirror. When the anti-Stokes sideband is resonant with the cavity ($\Delta =\protect\omega_{m}$),
an effective flow of energy from the mirror out of the cavity leads to an effective cooling.} \label{detuning_scheme}
\end{figure}

In analogy with well-known methods of atom and ion cooling \cite{Stenholm86,Leibfried03}, one can also think of cooling the mechanical resonator
by exploiting its coherent coupling to a fast decaying system which provides an additional dissipation channel and thus cooling. In the present
situation, radiation pressure couples the resonator with the cavity mode and the fast decaying channel is provided by the cavity photon loss
rate $\kappa $. An equivalent description of the process can be given in terms of dynamical backaction \cite{bragbook,brag}: the cavity reacts
with a delay to the mirror motion and induces correlations between the radiation pressure force and the Brownian motion that lead to cooling or
amplification, depending on the laser detuning. A quantitative description is provided by considering scattering of laser photons into the
motional sidebands induced by the mirror motion (see Fig.~\ref{detuning_scheme}) \cite{marquardt,wilson-rae,genes07}. Stokes (red) and
anti-Stokes (blue) sidebands are generated in the cavity at frequencies $\omega _{l}\pm \omega _{m}$. Laser photons are scattered by the moving
oscillator into the two sidebands with rates
\begin{equation}
A_{\pm }=\frac{G^{2}\kappa }{2\left[ \kappa ^{2}+\left( \Delta \pm \omega _{m}\right) ^{2}\right] },  \label{rates}
\end{equation}%
simultaneously with the absorption (Stokes, $A_{+}$) or emission (anti-Stokes, $A_{-}$) of vibrational phonons. The inequality $A_{-}>A_{+}$
leads to a decrease in the oscillator phonon occupation number and thus to cooling. Eq.~(\ref{rates}) shows that this occurs when $\Delta >0$
and that an effective optical cooling rate,
\begin{equation}
\Gamma =A_{-}-A_{+}=\frac{2G^{2}\Delta \omega _{m}\kappa }{\left[ \kappa ^{2}+(\omega _{m}-\Delta )^{2}\right] \left[ \kappa ^{2}+(\omega
_{m}+\Delta )^{2}\right] },  \label{gamma}
\end{equation}
can be defined, providing a measure of the coupling rate of the resonator with the effective zero-temperature environment represented by the
decaying cavity mode. Since the mechanical damping rate $\gamma _{m}$ is the coupling rate with the thermal reservoir of the resonator, one can
already estimate that, when $\Gamma \gg \gamma _{m}$, the mechanical oscillator is cooled at the new temperature $T\simeq \left( \gamma
_{m}/\Gamma \right) T_{0}$.

\begin{figure}[tb]
\centerline{\includegraphics[width=0.95\textwidth]{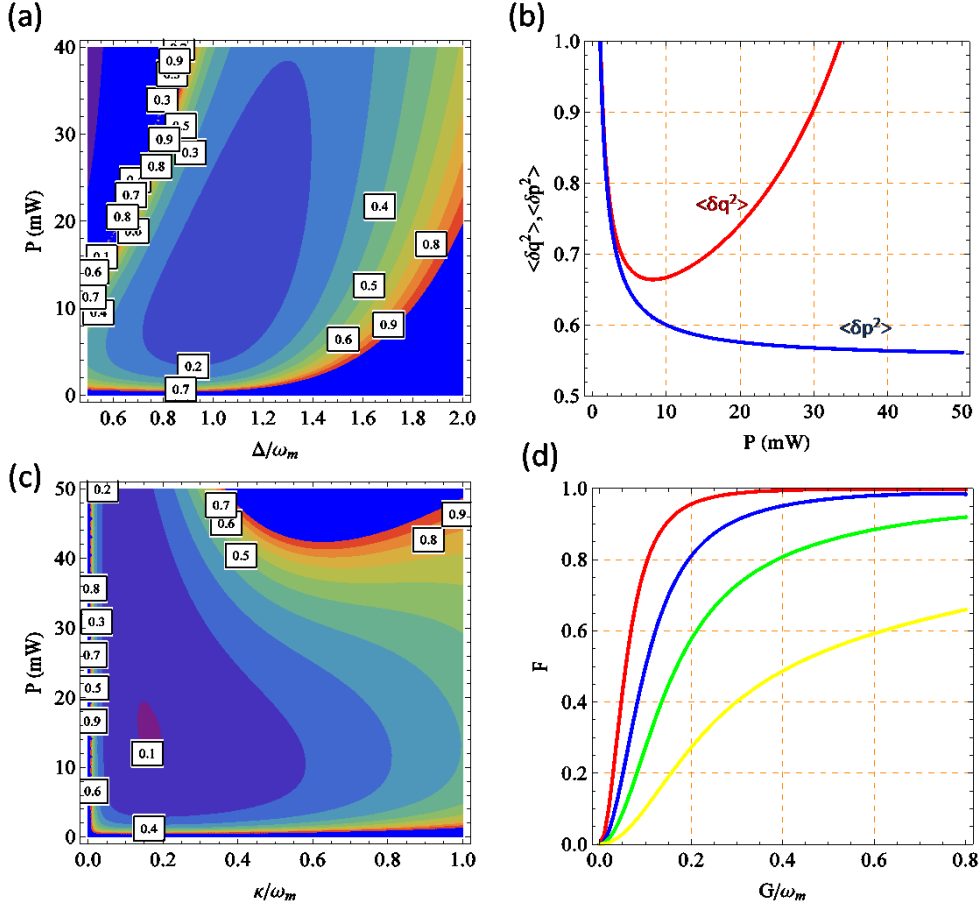}} \caption{Back action cooling. (a) Contour plot of $n$
versus $\Delta /\protect\omega _{m}$ and $\mathcal{P}.$ The parameters are $p_{0}$ and $\protect\kappa =0.37\protect\omega _{m}$. Optimal
cooling is seen to emerge around $\Delta = \protect\omega _{m}$. (b) For large $\mathcal{P}$ extra shot-noise is fed-back into the position
variance and the mirror thermalizes in a state where the equipartition theorem does not hold. (c) Contour plot of $n$ vs. $\protect\kappa
/\protect\omega _{m}$ and $\mathcal{P}$ for $\Delta =\protect\omega _{m}$. Optimal cooling is achieved around $\protect\kappa \simeq 0.2\times
\protect\omega _{m}$. (d) Fidelity between the mirror and intracavity states in the cooling regime as a function
of increasing intensity $G/\protect\omega _{m}$ with different values of $%
\protect\kappa /\protect\omega _{m}=0.2$ (red line), $0.5$ (blue), $1$ (green) and $2$ (yellow).}\label{cooling_detuning}
\end{figure}

One can perform a more precise and rigorous derivation of the cooling rate and steady state occupancy by using Eq.~(\ref{meanenergy}). The
position and momentum variances can be in fact obtained by solving Eq.~(\ref{Lyapunov}) or, equivalently, by solving the linearized QLE in the
Fourier domain and integrating the resulting noise spectra. The result of these calculations, in the limit of large mechanical quality factor
$\mathcal{Q}_m$, reads
\begin{eqnarray}
\left\langle \delta p^{2}\right\rangle & = &\frac{1}{\gamma _{m}+\Gamma }
\left\{ \frac{A_{+}+A_{-}}{2}+\gamma _{m}n_{0}\left( 1+\frac{\Gamma }{%
2\kappa }\right) \right\} , \label{qandp1}\\
\left\langle \delta q^{2}\right\rangle & = & \frac{1}{\gamma _{m}+\Gamma }%
\left\{ a\frac{A_{+}+A_{-}}{2}+\frac{\gamma _{m}n_{0}}{\eta }\left( 1+\frac{%
\Gamma }{2\kappa }b\right) \right\}, \label{qandp2}
\end{eqnarray}
where $\eta $ is given by Eq.~(\ref{eta}),
\begin{eqnarray}
a& =& \frac{\kappa ^{2}+\Delta ^{2}+\eta \omega _{m}^{2}}{\eta \left( \kappa
^{2}+\Delta ^{2}+\omega _{m}^{2}\right) }, \\
b& = & \frac{2\left( \Delta ^{2}-\kappa ^{2}\right) -\omega _{m}^{2}}{\kappa ^{2}+\Delta ^{2}}.
\end{eqnarray}
In the perturbative limit $\omega _{m}\gg n_{0}\gamma _{m},G$ and $\kappa \gg \gamma _{m},G$, Eqs.~(\ref{qandp1})-(\ref{qandp2}) simplify to
$\left\langle \delta q^{2}\right\rangle \simeq \left\langle \delta p^{2}\right\rangle\simeq n+1/2$, with $n\simeq \left[\gamma
_{m}n_{0}+A_{+}\right]/\left[\gamma_m+\Gamma\right] $, which reproduces the result of \cite{marquardt,wilson-rae}. This indicates that ground
state cooling is reachable when $\gamma _{m}n_{0}<\Gamma $ and provided that the radiation pressure noise contribution $A_{+}/\Gamma \simeq
\kappa ^{2}/\left( 4\omega _{m}^{2}\right) $ is also small. The optical damping rate $\Gamma $ can be increased by cranking up the input cavity
power and thus $G$. However, when one considers the limitations imposed by the stability condition $\eta >0$, one finds that there is an upper
bound for $G$ and consequently $\Gamma $. This is shown in Figs.~\ref{cooling_detuning}a-\ref{cooling_detuning}c, where one sees that for the
chosen parameter regime $p_0$, optimal cooling is achieved for $\Delta \simeq \omega_m$ (when the anti-Stokes sideband is resonant with the
cavity, as expected), and in a moderate good-cavity condition, $\kappa/\omega_m \simeq 0.2$. Fig.~\ref{cooling_detuning}b shows that close to
this optimal cooling condition, equipartition is soon violated when the input power (and therefore the effective coupling $G$) is further
increased: the position variance becomes much larger than the momentum variance and it is divergent at the bistability threshold (see
Eq.~(\ref{qandp2})).

\subsection{Readout of the mechanical resonator state} \label{readout}

Eq.~(\ref{fourierfeed}) shows that the cavity output is sensitive to the resonator position. Therefore, after an appropriate calibration, the
cavity output noise power spectrum provides a direct measurement of the position noise spectrum $S_q(\omega)$ which, when integrated over
$\omega$, yields the value of the position variance (see Eq.~(\ref{spectra})). In many
experiments~\cite{arcizet06,gigan06,arcizet06b,bouwm,vahalacool,mavalvala,rugar,wineland,markusepl,harris}, this value is employed to estimate
the final effective temperature of the cooled resonator by assuming energy equipartition $\left\langle \delta p^{2}\right\rangle \simeq
\left\langle \delta q^{2}\right\rangle$ so that $n \simeq \left\langle \delta q^{2}\right\rangle -1/2$. However, as we have seen above (see
Eqs.~(\ref{qusqcd2}), (\ref{pisqcd2}), (\ref{qandp1}), (\ref{qandp2})), equipartition does not generally hold and one should rather estimate
$\left\langle \delta p^{2}\right\rangle $ from $S_q(\omega)$ using Eq.~(\ref{spectra}), or directly measure independently the resonator
momentum. A different and more direct way of measuring the resonator temperature, borrowed from trapped-ion experiments \cite{Leibfried03}, has
been demonstrated in \cite{sidebcooling}: if the two motional sidebands are well resolved and detected via heterodyne measurement, the height of
the two sideband peaks is proportional to $n$ and to $n+1$, so that one can directly measure the occupancy $n$ from the comparison of the two
peaks.

However, one should devise a scheme capable of reconstructing the \emph{complete quantum state} of the resonator from the cavity output light,
which is the only accessible degree of freedom carrying out information about it. In fact, the full reconstruction of the quantum state of the
resonator is a necessary condition for the unambiguous demonstration of the quantum behavior of the mechanical resonator, as for example
stationary entanglement, which will be discussed in the following. A scheme of this kind has been proposed in \cite{prl07}, based on the
transfer of the resonator state onto the output field of an additional, fast-decaying, ``probe'' cavity mode. In fact, the annihilation operator
of this probe cavity mode, $a_p$, obeys an equation analogous to the linearization of Eq.~(\ref{mode}),
\begin{equation}
\label{probe2} \delta \dot{a}_p =-(\kappa_p+i\Delta_p)\delta a_p +i G_p \alpha_p \delta q +\sqrt{2\kappa_2} a_{p}^{in}(t),
\end{equation}
where $\kappa_p$, $\Delta_p$, $G_p$, $\alpha_p$, and $a_{p}^{in}(t)$ are respectively the decay rate, the effective detuning, the coupling, the
intracavity field amplitude, and the input noise of the probe cavity mode. The presence of the probe mode affects the system dynamics, but if
the driving of the probe mode is much weaker so that $|\alpha_p| \ll |\alpha_s|$, the back-action of the probe mode on the resonator can be
neglected.
If one chooses parameters so that $\Delta_p=\omega_m \gg k_p, G_p |\alpha_p|$, one can rewrite Eq.~(\ref{probe2}) in the frame rotating at
$\Delta_p=\omega_m$ for the slow variables $\delta \tilde{o}(t) \equiv \delta o(t)\exp\{i\omega_m t\}$ and neglect fast oscillating terms, so to
get
\begin{equation}
\label{probe3} \delta \dot{\tilde{a}}_p=-\kappa_p\delta \tilde{a}_p +i \frac{G_p \alpha_p }{\sqrt{2}} \delta \tilde{b} +\sqrt{2\kappa_p}
\tilde{a}_{p}^{in}(t),
\end{equation}
where $\delta{b}=(i \delta p + \delta q)/\sqrt{2}$ is the mechanical annihilation operator. Finally, if $\kappa_p \gg G_p |\alpha_p| /\sqrt{2}$,
the probe mode adiabatically follows the resonator dynamics and one has
\begin{equation}
\label{probe4} \delta \tilde{a}_p \simeq i \frac{G_p \alpha_p }{\kappa_p\sqrt{2}} \delta \tilde{b} +\sqrt{\frac{2}{\kappa_p}}
\tilde{a}_{p}^{in}(t).
\end{equation}
The input-output relation $\tilde{a}_p^{out}=\sqrt{2\kappa_p}\delta \tilde{a}_p-\tilde{a}_{p}^{in}$ \cite{gard} implies
\begin{equation}
\label{output} \tilde{a}_p^{out}= i \frac{G_p \alpha_p }{\sqrt{\kappa_p}} \delta \tilde{b} + \tilde{a}_{p}^{in}(t),
\end{equation}
showing that, in the chosen parameter regime, the output light of the probe mode gives a direct measurement of the resonator dynamics. With an
appropriate calibration and applying standard quantum tomographic techniques \cite{tomo} to this output field, one can therefore reconstruct the
quantum state of the resonator.

An alternative way to detect the resonator state by means of state transfer onto an optical mode, which does not require an additional probe
mode, can be devised by appropriately exploiting the strong coupling regime. In this second example state transfer is realized in a transient
regime soon after the preparation of the desired resonator state. One sets the cavity onto resonance $\Delta =0$ so that the system is always
stable, and then strongly increases the input power in order to make the coupling $G$ very large, $G \gg \kappa,n_{0}\gamma _{m}$. Under these
conditions, coherent evolution driven by
radiation pressure dominates and one has state swapping from the mechanical resonator onto the intracavity mode in a time $%
t_{swap} \simeq \pi /2G$ so that the cavity mode state reproduces the resonator state with a fidelity very close to unity. The fidelity of the
swap can be computed and reads
\begin{eqnarray}
 F &=& \left[\sqrt{\det \left( \mathcal{V}_{1}+\mathcal{V}_{2}\right) +\left( \det \mathcal{V}_{1}-1/4\right) \left( \det
\mathcal{V}_{2}-1/4\right) }\right. \nonumber \\
& -& \left.  \sqrt{\left( \det \mathcal{V}_{1}-1/4\right) \left( \det \mathcal{V}%
_{2}-1/4\right) }\right]^{-1},
\end{eqnarray}
where $\mathcal{V}_{1},\mathcal{V}_{2}$ are the block matrices in Eq. (\ref%
{CMatrix}). The resulting fidelity under realistic conditions is plotted in Fig.~(\ref{cooling_detuning}d) as a function of $G/\omega _{m}$ for
$\kappa /\omega _{m}=0.2\,$, $0.5$, $1$ and $2$. One can see that the fidelity is close to unity around the optimal cooling regime and that in
this regime \textit{both the mechanical resonator and intracavity field thermalize in the same state.} Under this condition one can reconstruct
the quantum state of the mechanical mode from the detection of the cavity output.

\section{Entanglement generation with a single driven cavity mode} \label{singledriven}

As discussed in the introduction, a cavity coupled to a mechanical degree of freedom is capable of producing entanglement between the mechanical
and the optical modes and also purely optical entanglement between the induced motional sidebands. In the following we elucidate the physical
origins of this entanglement and analyze its magnitude and temperature robustness. Moreover, we analyze its use as a quantum-communication
network resource in which the mechanical modes play the role of local nodes that store quantum information and optical modes carry this
information among nodes. To this purpose we apply a multiplexing approach that allows one, by means of spectral filters, to select many traveling
output modes originating from a single intracavity field.

\subsection{Intracavity optomechanical entanglement}

Entanglement can be easily evaluated and quantified using the logarithmic negativity of Eq.~(\ref{eq:logneg}), which requires the knowledge of
the CM of the system of interest. For the steady state of the intracavity field-resonator system, the CM is determined in a straightforward way
by the solution of Eq.~(\ref{Lyapunov}). However, before discussing the general result we try to give an intuitive idea of how robust
optomechanical entanglement can be generated, by using the sideband picture. Using the mechanical annihilation operator $\delta b$ introduced in
the above section, the linearized QLE of Eqs.~(\ref{lle1})-(\ref{lle4}) can be rewritten as
\begin{eqnarray}
\delta \dot{\tilde{b}}& =& -\frac{\gamma _{m}}{2}\left( \delta \tilde{b}%
-\delta \tilde{b}^{\dagger }e^{2i\omega _{m}t}\right) +\sqrt{\gamma _{m}}%
b^{in}+i\frac{G}{2}\left( \delta \tilde{a}^{\dag }e^{i(\Delta +\omega _{m})t}+\delta \tilde{a}e^{i(\omega _{m}-\Delta )t}\right) ,
\label{mode2}
\\
\delta \dot{\tilde{a}}& =&-\kappa \delta \tilde{a}+i\frac{G}{2}\left( \delta \tilde{b}^{\dag }e^{i(\Delta +\omega _{m})t}+\delta
\tilde{b}e^{i(\Delta -\omega _{m})t}\right) +\sqrt{2\kappa }\tilde{a}^{in}.  \label{modemech2}
\end{eqnarray}%
We have introduced the tilded slowly evolving operators $\delta \tilde{b}(t)=\delta b(t)e^{i\omega _{m}t}$, $\delta \tilde{a}(t)=\delta
a(t)e^{i\Delta t}$, and the noises $\tilde{a}^{in}(t)=a^{in}(t)e^{i\Delta t}$ and $b^{in}(t)=\xi (t)e^{i\omega _{m}t}/\sqrt{2}$. The input noise
$\tilde{a}^{in}(t)$ possesses the same correlation function as $a^{in}(t)$, while the Brownian noise $ b^{in}(t)$ in the limit of large
mechanical frequency $\omega_m$ acquires ``optical-like'' correlation functions $\langle
b^{in,\dag }(t)b^{in}(t^{\prime })\rangle =n_{0}\delta (t-t^{\prime })$ and $%
\langle b^{in}(t)b^{in,\dag }(t^{\prime })\rangle =\left[ n_{0}+1\right] \delta (t-t^{\prime })$ \cite{gard2}.
Eqs.~(\ref{mode2})-(\ref{modemech2}) show that the cavity mode and mechanical resonator are coupled by radiation pressure via two kinds of
interactions: i) a down-conversion process with interaction Hamiltonian $\delta \tilde{b}^{\dag }\delta \tilde{a}^{\dag }+\delta \tilde{a}\delta
\tilde{b}$, which is modulated by a factor oscillating at $\omega_m + \Delta$;  ii) a beam-splitter-like process with interaction Hamiltonian
$\delta \tilde{b}^{\dagger }\delta \tilde{a}+\delta \tilde{a}^{\dagger }\delta \tilde{b}$, modulated by a factor oscillating at $\omega_m -
\Delta$. Therefore, by tuning the cavity into resonance with either the Stokes sideband of the driving laser, $\Delta =-\omega _{m}$, or the
anti-Stokes sideband of the driving laser, $\Delta =\omega _{m}$, one can resonantly enhance one of the two processes. In the rotating wave
approximation (RWA), which is justified in the limit of $\omega _{m}\gg G,\kappa $, the off-resonant interaction oscillates very fast with
respect to the timescales of interest and can be neglected. Therefore, in the RWA regime, when one chooses $\Delta =-\omega _{m}$, the radiation
pressure induces a down-conversion process, which is known to generate bipartite CV entanglement. Instead when one chooses $\Delta =\omega
_{m}$, the dominant process is the beam-splitter-like interaction, which is not able to generate optomechanical entanglement starting from
classical input states \cite{kim}, as in this case. This argument leads to the conclusion that, in the RWA limit $\omega _{m}\gg G,\kappa $, the
best regime for optomechanical entanglement is when the laser is blue-detuned from the cavity resonance $\Delta =-\omega _{m} $ and
down-conversion is enhanced. However, this argument is valid only in the RWA limit and it is strongly limited by the stability conditions, which
rather force to work in the opposite regime of a red-detuned laser. In fact, the stability condition of Eq.~(\ref{stab1}) in the RWA limit
$\Delta =-\omega _{m}\gg \kappa ,\gamma _{m} $, simplifies to $G<\sqrt{2\kappa \gamma _{m}}$. Since one needs small mechanical dissipation rate
$\gamma _{m}$ in order to see quantum effects, this means a very low maximum value for $G$. The logarithmic negativity $E_{\mathcal{N}}$ is an
increasing function of the effective optomechanical coupling $G$ (as expected) and therefore the stability condition puts a strong upper bound
also on $E_{\mathcal{N}}$. It is possible to prove that the following bound on $E_{\mathcal{N}}$ exists \cite{output08}
\begin{equation}
E_{\mathcal{N}}\leq \ln \left[ \frac{1+G/\sqrt{2\kappa \gamma _{m}}}{1+n_{0}}%
\right] ,  \label{logneg2}
\end{equation}%
showing that $E_{\mathcal{N}}\leq \ln 2$ and above all that entanglement is extremely fragile with respect to temperature in the blue-detuned
case because, due to the stability constraints, $E_{\mathcal{N}}$ vanishes as soon as $n_{0}\geq 1$.

\begin{figure}[tb]
\centerline{\includegraphics[width=0.95\textwidth]{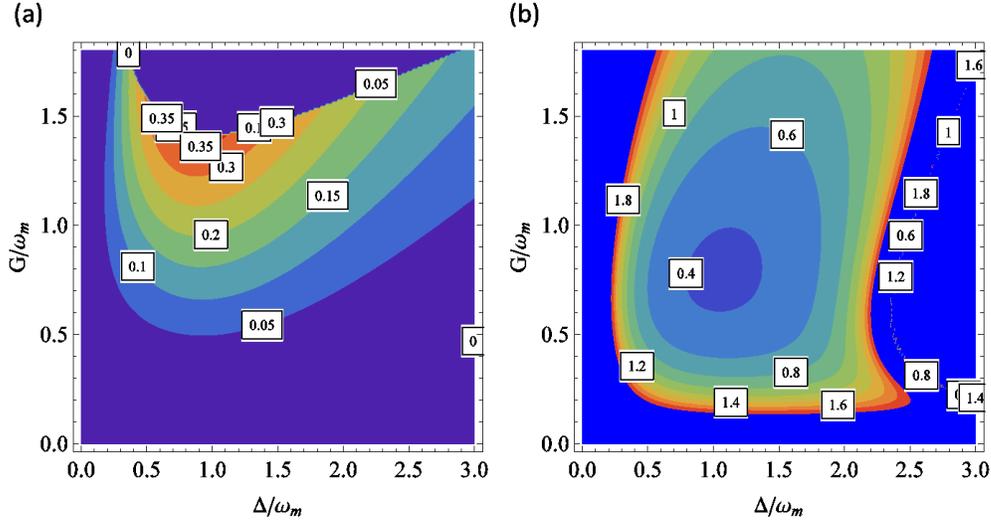}} \caption{Intracavity entanglement and cooling in the
red-detuned regime. (a) Contour plot of logarithmic
negativity of the field-mirror system at the steady state as a function of $G/\protect\omega %
_{m}$ and $\Delta /\protect\omega _{m}$ for the parameters $p_{0}$ and $\protect\kappa =%
\protect\omega _{m}$. (b) $n$ in the same parameter region: the plot shows that optimal cooling and optimal entanglement are both achieved close
to $\Delta /\protect\omega _{m}\simeq 1$. However, optimal cooling is obtained for smaller values of $G/\protect\omega_{m}$ with respect to
entanglement.}\label{intracavity_ent}
\end{figure}

This suggests that, due to instability, one can find significant intracavity optomechanical entanglement, which is also robust against
temperature, only far from the RWA regime, in the strong coupling regime in the region with positive $\Delta $, because Eq.~(\ref{stab2}) allows
for higher values of the coupling ($G< \sqrt{\kappa^2+\omega_m^2}$ when $\Delta=\omega_m$). This is confirmed by Fig.~\ref{intracavity_ent}a,
where the \emph{exact} $E_{\mathcal{N}}$ calculated from the solution of Eq.~(\ref{Lyapunov}) is plotted versus the normalized detuning $\Delta
/\omega _{m}$ and the normalized effective optomechanical coupling $G/\omega _{m}$. One sees that $E_{\mathcal{N}}$ reaches significant values
close to the bistability threshold; moreover it is possible to see that such intracavity entanglement is robust against thermal noise because it
survives up to reservoir temperatures around $20$ K \cite{prl07}. It is also interesting to compare the conditions for optimal entanglement and
cooling in this regime where the cavity is resonant with the anti-Stokes sideband. In Fig.~\ref{intracavity_ent}b, $n$ is plotted versus the
same variables in the same parameter region. One can see that, while good entanglement is accompanied by good cooling, optimal entanglement is
achieved for the largest possible coupling $G$ allowed by the stability condition. This condition is far from the optimal cooling regime, which
does not require very large $G$ because otherwise the radiation pressure noise contribution and consequently the position variance become too
large (see Eq.~(\ref{qandp2}) and Fig.~\ref{cooling_detuning}) \cite{output08}.

\subsection{Entanglement with output modes}\label{outputform}

Let us now define and evaluate the entanglement of the mechanical resonator with the fields at the cavity output, which may represent an
essential tool for the future integration of micromechanical resonators as quantum memories within quantum information networks. The intracavity
field $\delta a(t)$ and its output are related by the usual input-output relation \cite{gard}
\begin{equation}
a^{out}(t)=\sqrt{2\kappa}\delta a(t)-a^{in}(t),  \label{inout}
\end{equation}
where the output field possesses the same correlation functions of the optical input field $a^{in}(t)$ and the same commutation relation, i.e.,
the
only nonzero commutator is $\left[ a^{out}(t),a^{out}(t^{\prime})^{\dagger }%
\right] =\delta(t-t^{\prime})$. From the continuous output field $a^{out}(t)$ one can extract many independent optical modes, by selecting
different time
intervals or equivalently, different frequency intervals (see e.g. \cite%
{fuchs}). One can define a generic set of $N$ output modes by means of the corresponding annihilation operators
\begin{equation}
a_{k}^{out}(t)=\int_{-\infty}^{t}dsg_{k}(t-s)a^{out}(s),\;\;\;k=1,\ldots N, \label{filter1}
\end{equation}
where $g_{k}(s)$ is the causal filter function defining the $k$-th output mode. These annihilation operators describe $N$ independent optical
modes when $\left[ a_{j}^{out}(t),a_{k}^{out}(t)^{\dagger}\right] =\delta_{jk}$, which is verified when
\begin{equation}
\int_{0}^{\infty}dsg_{j}(s)^{\ast}g_{k}(s)=\delta_{jk},  \label{filter2}
\end{equation}
i.e., the $N$ filter functions $g_{k}(t)$ form an orthonormal set of square-integrable functions in $[0,\infty)$. The situation can be
equivalently described in the frequency domain: taking the Fourier transform of Eq.~(\ref{filter1}), one has
\begin{equation}
\tilde{a}_{k}^{out}(\omega)=\int_{-\infty}^{\infty}\frac{dt}{\sqrt{2\pi}}%
a_{k}^{out}(t)e^{i\omega t}=\sqrt{2\pi}\tilde{g}_{k}(\omega)a^{out}(\omega), \label{filterFT}
\end{equation}
where $\tilde{g}_{k}(\omega)$ is the Fourier transform of the filter
function. 
An explicit example of an orthonormal set of filter functions is given by
\begin{equation}
g_{k}(t)=\frac{\theta(t)-\theta(t-\tau)}{\sqrt{\tau}}e^{-i\Omega_{k}t}, \label{filterex}
\end{equation}
($\theta$ denotes the Heavyside step function) provided that $\Omega_{k}$ and $\tau$ satisfy the condition
\begin{equation}
\Omega_{j}-\Omega_{k}=\frac{2\pi}{\tau}p,\;\;\;\mathrm{integer}\;\;p. \label{interfer}
\end{equation}
These functions describe a set of independent optical modes, each centered around the frequency $\Omega_{k}$ and with time duration $\tau$,
i.e., frequency bandwidth $\sim1/\tau$, since
\begin{equation}
\tilde{g}_{k}(\omega)=\sqrt{\frac{\tau}{2\pi}}e^{i(\omega-\Omega_{k})\tau /2}%
\frac{\sin\left[ (\omega-\Omega_{k})\tau/2\right] }{(\omega-\Omega _{k})\tau/2}.  \label{filterex2}
\end{equation}
When the central frequencies differ by an integer multiple of $2\pi/\tau$, the corresponding modes are independent due to the destructive
interference of the oscillating parts of the spectrum.

The entanglement between the output modes defined above and the mechanical mode is fully determined by the corresponding $(2N+2)\times (2N+2)$
CM, which is defined by
\begin{equation}
\mathcal{V}_{ij}^{out}(t)=\frac{1}{2}\left\langle u_{i}^{out}(t)u_{j}^{out}(t)+u_{j}^{out}(t)u_{i}^{out}(t)\right\rangle , \label{defVout}
\end{equation}%
where
\begin{equation}
u^{out}(t)=\left( \delta q(t),\delta p(t),X_{1}^{out}(t),Y_{1}^{out}(t),\ldots ,X_{N}^{out}(t),Y_{N}^{out}(t)\right) ^{T}
\end{equation}%
is the vector formed by the mechanical position and momentum fluctuations
and by the amplitude ($X_{k}^{out}(t)=\left[ a_{k}^{out}(t)+a_{k}^{out}(t)^{%
\dagger }\right] /\sqrt{2}$), and phase ($Y_{k}^{out}(t)=\left[ a_{k}^{out}(t)-a_{k}^{out}(t)^{\dagger }\right] /i\sqrt{2})$ quadratures of the
$N$ output modes. The vector $u^{out}(t)$ properly describes $N+1$ independent CV bosonic modes, and in particular the mechanical resonator is
independent of (i.e., it commutes with) the $N$ optical output modes because the latter depend upon the output field at former times only
($s<t$). From the intracavity CM and Eqs.~(\ref{inout}),(\ref{filter1}), and (\ref{defVout}) one
can determine the $\left( N+1\right) \times \left( N+1\right) $ CM matrix $%
\mathcal{V}^{out}$ at the steady state \cite{output08}.

Let us first consider the case when we select and detect only one mode at the cavity output. Just to fix the ideas, we choose the mode specified
by the filter function of Eqs.~(\ref{filterex}) and (\ref{filterex2}), with central frequency $\Omega $ and bandwidth $\tau ^{-1}$.
Straightforward choices for this output mode are a mode centered either at the cavity frequency, $\Omega =\omega _{c}-\omega _{l}$, or at the
driving laser frequency, $\Omega =0$ (we are in the rotating frame and therefore all frequencies are referred to the laser frequency $\omega
_{l}$), and with a bandwidth of the order of the cavity bandwidth $\tau ^{-1}\simeq \kappa $. However, as discussed above, the motion of the
mechanical resonator generates Stokes and anti-Stokes motional sidebands, consequently modifying the cavity output spectrum.

\begin{figure}[tb]
\centerline{\includegraphics[width=0.95\textwidth]{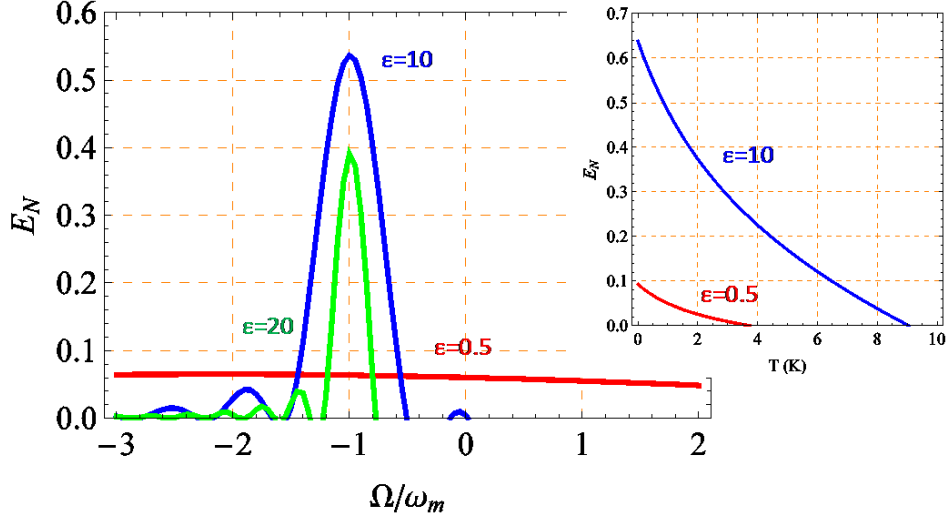}} \caption{Resonator-output field entanglement when the central
frequency of the output mode is swept around the laser frequency. Parameters are $p_{0}$, $\Delta=\omega_m$, $G=\omega_m/2$
and $\protect\kappa =\protect\omega
_{m}$.
The entanglement is optimized when the output mode coincides with the Stokes sideband of the laser ($\Omega =-\protect%
\omega _{m}$), with the appropriate bandwidth ($\protect\epsilon \simeq 10$, corresponding to $\tau \gamma _{m}^{eff}\simeq 1$). For smaller
$\protect\epsilon $, the selected output mode mixes Stokes and anti-Stokes photons and the entanglement is weak, while for larger $\epsilon$
only a fraction of the sideband is selected and part of the quantum correlations are lost. In the inset the robustness of Stokes-mirror
$E_{\mathcal{N}}$ with respect to temperature is shown.}\label{Stokes_ent}
\end{figure}

In order to determine the output optical mode which is better entangled with
the mechanical resonator, we study the logarithmic negativity $E_{\mathcal{N}%
}$ associated with the output CM $\mathcal{V}^{out}$ (for $N=1$) as a
function of the central frequency of the mode $\Omega $ and its bandwidth $%
\tau ^{-1}$, at the same parameter region considered in the previous subsection, $p_0$ and $\Delta=\omega_m$, where intracavity entanglement is
optimal. The results are shown in Fig.~\ref{Stokes_ent}, where $E_{\mathcal{N}}$ is plotted versus $\Omega
/\omega _{m}$ at different values of $\varepsilon =\tau \omega _{m}$. If $%
\varepsilon \lesssim 1$, i.e., the bandwidth of the detected mode is larger than $\omega _{m}$, the detector does not resolve the motional
sidebands, and $E_{\mathcal{N}}$ has a value (roughly equal to that of the intracavity case) which does not essentially depend upon the central
frequency. For smaller bandwidths (larger $\varepsilon $), the sidebands are resolved by the detection and the role of the central frequency
becomes important. In particular $E_{\mathcal{N}}$ becomes highly peaked around the \emph{Stokes sideband} $\Omega =-\omega _{m}$, showing that
the optomechanical entanglement generated within the cavity is mostly carried by this lower frequency sideband. What is relevant is that the
optomechanical entanglement of the output mode is significantly larger than its intracavity counterpart and achieves its maximum value at the
optimal value $\varepsilon \simeq 10$, i.e., a detection bandwidth $\tau ^{-1}\simeq \omega _{m}/10$. This means that in practice, by
appropriately filtering the output light, one realizes an \emph{effective entanglement distillation} because the selected output mode is more
entangled with the mechanical resonator than the intracavity field.

The fact that the output mode which is most entangled with the mechanical resonator is the one centered around the Stokes sideband is also
consistent
with the physics of a previous model analyzed in \cite{prltelep}. In \cite%
{prltelep}, a free-space optomechanical model is discussed, where the entanglement between a vibrational mode of a perfectly reflecting
micro-mirror and the two first motional sidebands of an intense laser beam shined on the mirror is analyzed. Also in that case, the mechanical
mode is entangled only with the Stokes mode and it is not entangled with the anti-Stokes sideband.

One can also understand why the output mode optimally entangled with the mechanical mode has a finite bandwidth $\tau ^{-1}\simeq \omega
_{m}/10$ (for the chosen operating point). In fact, the optimal situation is achieved when the detected output mode overlaps as best as possible
with the Stokes peak in the spectrum, and therefore $\tau ^{-1}$ coincides with the width of the Stokes peak. This width is determined by the
effective damping rate of the mechanical resonator, $\gamma _{m}^{eff}=\gamma _{m}+\Gamma $, given by the sum of the intrinsic damping rate
$\gamma _{m}$ and the net laser cooling rate $\Gamma $ of Eq.~(\ref{gamma}). It is possible to check that, with the chosen parameter values, the
condition $\varepsilon =10$ corresponds to $\tau ^{-1}\simeq \gamma _{m}^{eff}$.

It is finally important to analyze the robustness of the present optomechanical entanglement with respect to temperature. As discussed above and
shown in \cite{prl07}, the entanglement of the resonator with the intracavity mode is very robust. It is important to see if this robustness is
kept also by the optomechanical entanglement of the output mode. This is shown also in the inset of Fig.~\ref{Stokes_ent},
where the logarithmic negativity
$E_{\mathcal{N}}$ of the output mode centered at the Stokes sideband $\Omega =-\omega _{m}$ is plotted versus the temperature of the reservoir
at two different values of
the bandwidth, the optimal one $\varepsilon =10$, and at a larger bandwidth $%
\varepsilon =0.5$. We see the expected decay of $E_{\mathcal{N}}$ for increasing temperature, but above all that also this output optomechanical
entanglement is robust against temperature because it persists even above liquid He temperatures, at least in the case of the optimal detection
bandwidth $\varepsilon =10$.

\subsection{Optical entanglement between sidebands}

Let us now consider the case where we detect at the output two independent, well resolved, optical output modes. We use again the step-like
filter functions of Eqs.~(\ref{filterex}) and (\ref{filterex2}), assuming the same bandwidth $\tau ^{-1}$ for both modes and two different
central frequencies, $\Omega _{1}$ and $\Omega _{2}$, satisfying the orthogonality condition of Eq.~(\ref{interfer}) $\Omega _{1}-\Omega
_{2}=2p\pi \tau ^{-1}$ for some integer $p$, in order to have two independent optical modes. It is interesting to analyze the stationary state
of the resulting tripartite CV system formed by the two output modes and the mechanical mode, in order to see if and when it is able to show
purely optical bipartite entanglement between the two output modes.

The generation of two entangled light beams by means of the radiation pressure interaction of these fields with a mechanical element has been
already considered in various configurations. In Ref.~\cite{giovaEPL01}, and more recently in Ref.~\cite{wipf}, two modes of a Fabry-Perot
cavity system with a movable mirror, each driven by an intense laser, are entangled at the output due to their common ponderomotive interaction
with the movable mirror
(the scheme has been then generalized to many driven modes in \cite%
{giannini03}). In the single mirror free-space model of Ref.~\cite{prltelep}%
, the two first motional sidebands are also robustly entangled by the radiation pressure interaction as in a two-mode squeezed state produced by
a non-degenerate parametric amplifier \cite{jopbpir}. Robust two-mode squeezing of a bimodal cavity system can be similarly produced if the
movable mirror is replaced by a single ion trapped within the cavity \cite{morigi}.

\begin{figure}[tb]
\centerline{\includegraphics[width=0.95\textwidth]{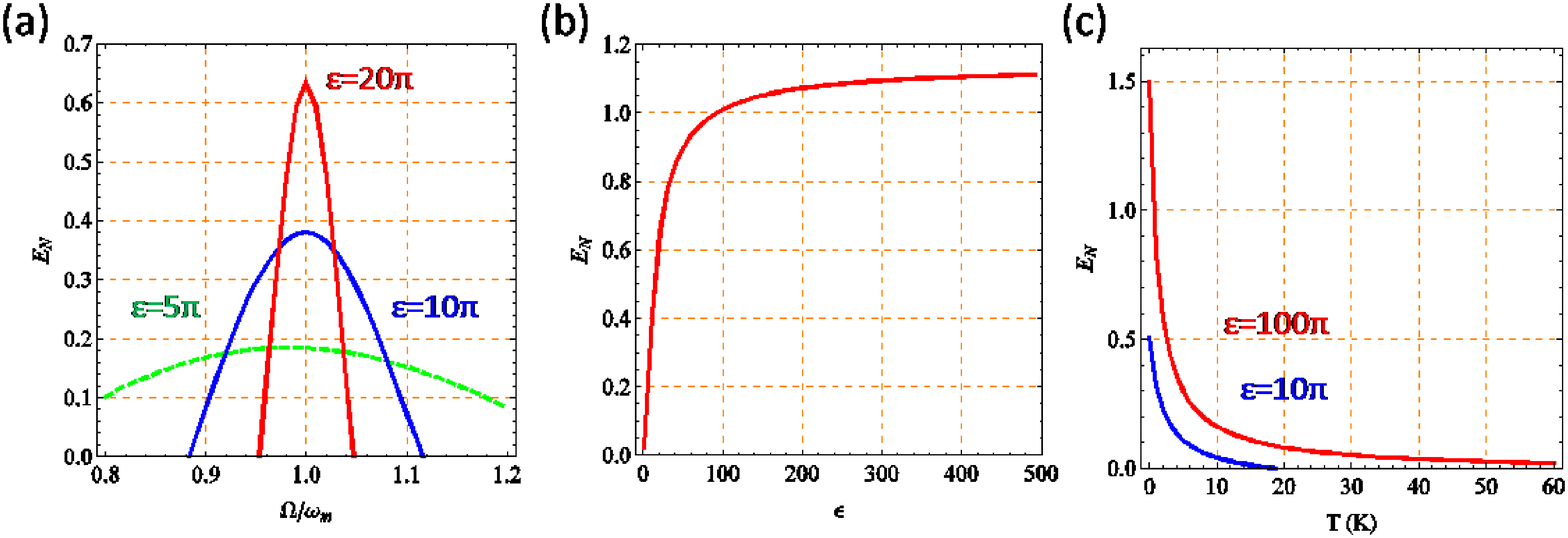}} \caption{Sideband-sideband entanglement.
Parameters $p_{0}$, $\protect\kappa =\protect\omega _{m}$ and $G=\protect%
\omega _{m}/2$. (a) Assuming one detection setup centered at the Stokes sideband and sweeping the second detection frequency around the
anti-Stokes sideband at $\Omega =\protect\omega _{m}$, the entanglement is clearly shown to be optimized when the anti-Stokes output field is
detected. This
entanglement is improving with smaller and smaller detection bandwidth ($%
\protect\epsilon \rightarrow \infty $). (b) Logarithmic negativity increases
asymptotically to a finite value with $\protect\epsilon \rightarrow \infty $%
. (c) Temperature robustness for $\protect\epsilon =10\protect\pi $ and $%
\protect\epsilon =100\protect\pi $. The entanglement survives to very high temperatures.}
\label{Sidebands_ent}
\end{figure}

The situation considered here is significantly different from that of Refs.~%
\cite{wipf,giovaEPL01,giannini03,morigi}, which require many driven cavity modes, each associated with the corresponding output mode. In the
present case instead, the different output modes \emph{originate from the same single driven cavity mode}, and therefore it is simpler from
an experimental point of view. The present scheme can be considered as a sort of \textquotedblleft cavity version\textquotedblright\ of the
free-space case of Ref.~\cite{prltelep}, where the reflecting mirror is driven by a single intense laser. Therefore, as in
\cite{prltelep,jopbpir}, one expects to find a parameter region where the two output modes centered around the two motional sidebands of the
laser are entangled. This expectation is clearly confirmed by Fig.~\ref{Sidebands_ent}a, where the logarithmic negativity $E_{\mathcal{N}}$
associated with the bipartite system formed by the output mode centered at the Stokes sideband ($\Omega _{1}=-\omega _{m}$) and a second output
mode with the same inverse bandwidth ($\varepsilon =\omega _{m}\tau =10\pi $) and a variable central frequency $\Omega $, is plotted versus
$\Omega /\omega _{m}$. $E_{\mathcal{N}}$ is calculated from the CM $\mathcal{V}^{out}$ (for $N=2$) eliminating the first two rows associated
with the mechanical mode. One can clearly see that bipartite entanglement between the two cavity outputs exists only in a narrow frequency
interval around the anti-Stokes sideband, $\Omega =\omega _{m}$,
where $E_{\mathcal{N}}$ achieves its maximum. This shows that, as in \cite%
{prltelep,jopbpir}, the two cavity output modes corresponding to the Stokes and anti-Stokes sidebands of the driving laser are significantly
entangled by their common interaction with the mechanical resonator. The advantage of
the present cavity scheme with respect to the free-space case of \cite%
{prltelep,jopbpir} is that the parameter regime for reaching radiation-pressure mediated optical entanglement is much more promising from an
experimental point of view because it requires less input power and a not
too large mechanical quality factor of the resonator. In Fig.~\ref%
{Sidebands_ent}b, the dependence of $E_{\mathcal{N}}$ of the two output modes centered at the two sidebands $\Omega =\pm \omega _{m}$ upon their
inverse bandwidth $\varepsilon $ is studied. We see that, differently from optomechanical entanglement of the former subsection, the logarithmic
negativity of the two sidebands always increases for decreasing bandwidth, and it achieves a significant value, comparable to that achievable
with parametric oscillators, for very narrow bandwidths. This fact can be understood from the fact that quantum correlations between the two
sidebands are established by the coherent scattering of the cavity photons by the oscillator, and that the quantum coherence between the two
scattering processes is maximal for output photons with frequencies $\omega _{l}\pm \omega _{m}$.
Figs.~\ref{Stokes_ent} and \ref{Sidebands_ent} show that in the chosen parameter regime,
the output mode centered around the Stokes sideband mode shows bipartite entanglement
simultaneously with the mechanical mode and with the anti-Stokes sideband mode. This fact suggests that the CV
tripartite system formed by the output Stokes and anti-Stokes sidebands and the mechanical resonator mode might be characterized by a fully
tripartite-entangled stationary state. This is actually true and it can be checked by
applying the classification criterion of
Ref.~\cite{giedke}, providing a necessary and sufficient criterion for the determination of the entanglement class in the case of tripartite CV
Gaussian states, which is directly computable in terms of the eigenvalues of appropriate test matrices \cite{giedke} (see Ref.~\cite{output08}).

\section{Entanglement generation with two driven cavity modes}

We now generalize the system by considering the case when \emph{two} cavity modes with different frequencies are intensely driven. We shall
focus onto a parameter regime which will prove to be convenient for the generation of robust stationary CV entanglement between the resonator and
the two cavity modes. A bichromatic driving of a cavity has been already experimentally considered in Refs.~\cite{mavalvala}. There however it
was employed for cooling a macroscopically heavy ($m\simeq 1 g$) movable mirror. One driven mode is used to ``trap'' the mirror, i.e., to induce
a strong optical spring effect increasing by three orders of magnitude the oscillation frequency. The other driven mode is instead used to cool
the mechanical resonator by increasing the effective mechanical damping, either via back-action, or via cold-damping feedback. The bichromatic
driving configuration has been already considered for the generation of entanglement in various configurations in some theoretical proposals. In
fact, in Ref.~\cite{giovaEPL01}, and more recently in Ref.~\cite{wipf}, two modes of a Fabry-Perot cavity system, each driven by an intense
laser, are entangled at the output due to their common ponderomotive interaction with the movable mirror.

\subsection{Quantum Langevin equations and stability conditions}

We generalize the Hamiltonian of Eq.~(\ref{Ham}) by considering \emph{two} cavity modes, with frequencies $\omega_{cA}$ and $\omega_{cB}$, each
driven by a laser with frequency $\omega_{0A}$ and $\omega_{0B}$, and power $\mathcal{P}_A$ and $\mathcal{P}_B$, respectively. The resulting Hamiltonian is
\begin{eqnarray}
&& H=\hbar\omega_{cA}\,a^{\dagger}a+\hbar\omega_{cB}\,b^{\dagger}b+\frac{1}{2}\hbar\omega_{m}(p^{2}+q^{2})
-\hbar( G_{0A}\,a^{\dagger}a + G_{0B}\,b^{\dagger}b) q  \notag \\
&& +i\hbar[ E_A(a^{\dagger}e^{-i\omega_{0A}t}-ae^{i\omega_{0A}t})+E_B(b^{\dagger}e^{-i\omega_{0B}t}-be^{i\omega_{0B}t})],  \label{ham0bis}
\end{eqnarray}
where $a$ and $b$ now denote the annihilation operators of the two cavity modes, we have introduced the two coupling constants
$G_{0x}=\sqrt{\hbar/m \omega_m}\,\omega_{cx}/L$, and the two driving rates $|E_x|=\sqrt{2P_x \kappa/\hbar \omega_{0x}}$, $x=A,B$. We have assumed for
simplicity that the two modes have the same decay rate $\kappa$. We assume that scattering of photons of the driven modes into
other cavity modes and also between the two chosen modes is negligible: this is verified when $\omega_m$ is much smaller then
the free spectral range of the cavity.

Introducing again dissipation and noise terms as in Sec.~\ref{sec1}, the system dynamics is described by the following set of nonlinear QLE,
written in the interaction picture with respect to $\hbar \omega_{0A} a^{\dag}a+\hbar \omega_{0B} b^{\dag}b$,
\begin{eqnarray}
\dot{q}&=&\omega_m p, \label{QLEam1}\\
\dot{p}&=&-\omega_m q - \gamma_m p + G_{0A} a^{\dag}a+G_{0B}b^\dag b + \xi, \label{QLEam2}\\
\dot{a}&=&-[\kappa+i(\Delta_{0A}-G_{0A}q)]a +E_A +\sqrt{2\kappa} a^{in},\label{QLEam3}\\
\dot{b}&=&-[\kappa+i(\Delta_{0B}-G_{0B}q)]b +E_B +\sqrt{2\kappa} b^{in}, \label{QLEam4}
\end{eqnarray}
where $\Delta_{0x}\equiv\omega_{cx}-\omega_{0x}$ are the detunings of the two lasers, and we have introduced a vacuum input noise $b^{in}(t)$
for the cavity mode $b$, possessing the same correlations of Eqs.~(\ref{input1})-(\ref{input2}).

We assume again that both modes are intensely driven so that the system is characterized by a semiclassical steady state with large intracavity
amplitudes
for both modes and a modified cavity length. This classical steady state is determined by setting the time derivatives
to zero, factorizing the averages and solving for the mean values $a_s=\langle a \rangle$, $b_s=\langle b \rangle$, $q_s=\langle q \rangle$,
$p_s=\langle p \rangle$. One gets
\begin{eqnarray}
&& a_s = \frac{E_A}{\kappa+i \Delta_A}, \label{as}  \\
&& b_s = \frac{E_B}{\kappa+i \Delta_B}, \label{bs}   \\
&& q_s = \frac{G_{0A} |a_s|^2 + G_{0B} |b_s|^2}{\omega_m}, \label{qs}\\
&& p_s = 0,
\end{eqnarray}
where the effective detunings $\Delta_{x} \equiv \Delta_{0x}- (G_{0A}^2 |a_s|^2+G_{0B}^2 |b_s|^2)/\omega_m$, $x=A,B$, have been defined, so that
Eqs.~(\ref{as})-(\ref{bs}) form actually a system of nonlinear equations, whose solution gives the stationary amplitudes $a_s$ and $b_s$.

One then focuses onto the dynamics of the quantum fluctuations around this steady state, which are well described by linearizing the QLE of
Eqs.~(\ref{QLEam1})-(\ref{QLEam4}) around the semiclassical steady state values, provided that $|a_s|, |b_s| \gg 1$. The linearized QLE
for the resonator and for the amplitude and phase quadratures of the two modes, $\delta X_A$, $\delta X_B$, $\delta Y_A$ and $\delta Y_B$,
defined as in Sec.~\ref{sec1}, can be written in compact form as
$$
\dot{u} (t) =A u(t)+ n(t), $$
where $ u = (\delta q,\delta p, \delta X_A,\delta Y_A,\delta X_B,\delta Y_B)^T$ is the vector of quadrature fluctuations,
and $ n = (0,\xi,\sqrt{2\kappa} X_A^{in},\sqrt{2\kappa} Y_A^{in},\sqrt{2\kappa} X_B^{in},\sqrt{2\kappa} Y_B^{in})^T$ is the corresponding vector of noises. The $6 \times 6$
matrix $A$ is the drift matrix of the system, which reads
\begin{equation}
 A=\left( \begin{array}{cccccc}
0    &\omega_m&0        &0       &0        &0   \\
-\omega_m&\gamma_m&G_A      &0       &G_B      &0   \\
0    &0   &-\kappa  &\Delta_A&0        &0   \\
G_A  &0   &-\Delta_A&-\kappa &0        &0   \\
0    &0   &0        &0       &-\kappa  &\Delta_B\\
G_B  &0   &0        &0       &-\Delta_B&-\kappa \\
 \end{array}\right),
\end{equation}
where we have chosen the phase reference of the two cavity modes so that $a_s$ and $b_s$ are real and positive, and defined the effective couplings
$G_A=G_{0A} a_s \sqrt{2}$ and $G_B=G_{0B} b_s \sqrt{2}$.

The steady state exists and it is stable if all the eigenvalues of the drift matrix $A$ have negative real parts.
The parameter region under which stability is verified can be obtained from the Routh-Hurwitz criteria \cite{grad}, but the inequalities that
come out in the present case are quite involved. One can have an idea of this fact from the expression of the
characteristic polynomial of $A$, $P(\lambda)=\lambda^6+c_1\lambda^5+c_2\lambda^4+c_3\lambda^3+c_4\lambda^2+c_5\lambda+c_6$, where
\begin{eqnarray}
c_1&=&\gamma_m+4\kappa, \notag \\
c_2&=& \Delta_A^2+\Delta_B^2+4 \gamma_m \kappa+6\kappa^2+\omega_m^2,\notag\\
c_3&=&\gamma_m (\Delta_A^2+\Delta_B^2+6 \kappa^2)+2 \kappa [\Delta_A^2+\Delta_B^2+2(\kappa^2+\Omega_m^2)], \notag \\
c_4&=& \kappa^4+2 \gamma_m \kappa (\Delta_B^2+2 \kappa^2)+6 \kappa^2 \omega_m^2+\Delta_B^2 (\kappa^2+\omega_m^2)+\notag\\
&&\Delta_A^2 (\Delta_B^2+2 \gamma_m \kappa +\kappa^2+\omega_m^2)-\omega_m(G_A^2 \Delta_A+G_B^2 \Delta_B), \notag\\
c_5&=& \gamma_m (\Delta_A^2+\kappa^2) (\Delta_B^2+\kappa^2)+2 \kappa\omega_m^2(\Delta_A^2+\Delta^2+2 \kappa^2)\notag\\
&& - 2\kappa\omega_m  (G_A^2 \Delta_A+G_B^2 \Delta_B),\notag\\
c_6&=&\omega_m^2 (\Delta_A^2+\kappa^2)(\Delta_B^2+\kappa^2)-\omega_m [G_B^2\Delta_B (\Delta_A^2+\kappa^2) \notag\\
&&  +G_A^2 \Delta_A(\Delta_B^2+\kappa^2)]. \notag
\end{eqnarray}
We are considering here a bichromatic driving of the cavity in order to improve the size and the robustness of the generated entanglement. Entanglement
monotonically increases with the optomechanical coupling but, as we have seen also in the previous sections, the stability conditions put a strict
upper bound on the maximum achievable value of this coupling. Therefore it is interesting to find a regime in which the presence of the
second driven mode makes the system always stable, so that the couplings can be made very large (for example by increasing the input power,
the cavity finesse, or decreasing the cavity length) without entering the unstable regime. One then hopes that in this regime also entanglement can be
made large and robust against temperature.

A simple way to have always a stable system is to find a particular relation between the parameters such that
the characteristic polynomial of $A$ does not depend upon $G_A$ and $G_B$.
In this case, the eigenvalues of $A$ would be independent of the two couplings and stability would be guaranteed.
The expressions above show that the eigenvalues of $A$ are independent of $G_A$ and $G_B$ and the system is always stable if and only if
\begin{subequations}\label{stabtwo}
\begin{eqnarray}
&&|G_a|=|G_B|=G, \\
&&\Delta_A=-\Delta_B=\Delta.
\end{eqnarray}
\end{subequations}
The condition described by Eqs.~(\ref{stabtwo}) represents a
perfect balance between a cooling cavity mode (which, without loss of generality, we can take as mode $A$, so that
$\Delta_A>0$) and a heating cavity mode, i.e., mode $B$ with $\Delta_B<0$. The fact that the eigenvalues of $A$ do not depend upon the couplings means
that the decay rates of both the resonator and the cavity modes are left unchanged and in this case radiation pressure mainly
create quantum correlations, i.e., entanglement, between the modes. We shall assume conditions (\ref{stabtwo}) from now on.

\subsection{Entanglement of the output modes}

We now calculate the entanglement properties of the steady state of the bichromatically driven cavity. However we shall not discuss here
the intracavity entanglement, but only the entanglement properties of the optical \emph{output modes}. In fact, as we have seen above
in the case of a single driven mode, one can obtain a larger optomechanical entanglement with respect to the intracavity case
by appropriately filtering the output modes.
Moreover only the entanglement with output modes is relevant for any quantum communication application. We shall apply therefore the filter function
formalism developed in Sec.~\ref{outputform}, restricted however here to the simple case of \emph{a single output mode} for each intracavity mode.
In fact, we have now two driven cavity modes and considering the more general case of multiple output modes associated to each driven mode
as in Sec.~\ref{outputform}, would render the description much more involved without however gaining too much insight into the physics of the problem.
The two output modes originate from two different cavity modes, and since the latter are not too close in frequency, they consequently describe
two independent modes. Therefore we do not need orthogonal filter functions like those of Eq.~(\ref{filterex})
used for the single driven mode case, and we choose here a different filter function. We consider the two output modes with
annihilation operators
\begin{equation}
 a_{\Omega_x}^{out}(t)=\int_{-\infty}^{t}ds g_x(t-s) a_x^{out}(s) \;\;\;x=A,B,
\end{equation}
where $a_A^{out}(t)$ and $a_B^{out}(t)$ are the usual output fields associated with the two cavity modes and
\begin{equation}
 g_x(t)=\sqrt{\frac{2}{\tau}} e^{-\left(1/\tau+i\Omega_x \right) t}\theta(t) \;\;\;x=A,B
\end{equation}
are the two filter functions, describing two output modes, both with bandwidth $1/\tau$ and with central frequencies, $\Omega_A$ and $\Omega_B$,
which are in general different from the cavity mode frequencies $\omega_{cA}$ and $\omega_{cB}$.

The entanglement between the chosen output modes and the mechanical resonator mode is fully determined by the corresponding $6\times 6$
CM, which is defined as in Eq.~(\ref{defVout})
\begin{equation}
\mathcal{V}_{ij}^{out}(t)=\frac{1}{2}\left\langle u_{i}^{out}(t)u_{j}^{out}(t)+u_{j}^{out}(t)u_{i}^{out}(t)\right\rangle , \label{defVout2}
\end{equation}%
where now
\begin{equation}
u^{out}(t)=\left[0,0,\delta X_{\Omega_A}^{out}(t),\delta Y_{\Omega_A}^{out}(t),\delta X_{\Omega_B}^{out}(t),\delta
Y_{\Omega_B}^{out}(t)\right]^T\,,
\end{equation}
is the vector formed by the mechanical position and momentum fluctuations and by the amplitude and phase quadratures of the filtered modes.
Using the various definitions, input-output relations and also the correlation function of the noise terms, one can derive an integral expression
for the CM $\mathcal{V}^{out}$ of the system (see Ref.~\cite{output08} for the details in a similar calculation), which is given by
\begin{equation}
\mathcal{V}^{out}=\int d\omega
\tilde{T}(\omega)\left[\tilde{M}(\omega)+\frac{P_{out}}{2\kappa}\right]D(\omega)\left[\tilde{M}(\omega)^{\dagger}+\frac{P_{out}}{2\kappa}
\right]\tilde{T}(\omega)^{\dagger}, \label{Vfinoutbis}
\end{equation}
where $\tilde T(\omega)$ is the Fourier transform of
\begin{small}
\begin{equation}\label{transform2}
  T(t)=\left(\begin{array}{cccccc}
    \delta(t) & 0 & 0 & 0 & 0 & 0 \\
     0 & \delta(t) & 0 & 0 & 0 & 0\\
    0 & 0 & \sqrt{2\kappa}{\rm Re}g_A(t) & -\sqrt{2\kappa}{\rm Im}g_A(t) & 0 & 0\\
    0 & 0 & \sqrt{2\kappa}{\rm Im}g_A(t) & \sqrt{2\kappa}{\rm Re}g_A(t)& 0 & 0 \\
    0 & 0 &  0 & 0 & \sqrt{2\kappa}{\rm Re}g_B(t) & -\sqrt{2\kappa}{\rm Im}g_B(t)\\
    0 & 0 &  0 & 0 & \sqrt{2\kappa}{\rm Im}g_B(t) & \sqrt{2\kappa}{\rm Re}g_B(t)\\
 \end{array}\right),
\end{equation}
\end{small},
\begin{equation}
\tilde{M}(\omega)=\left(i\omega+A\right)^{-1},
\end{equation}
$P_{out}={\rm Diag}[0,0,1,1,1,1]$ is the projector onto the optical quadratures,
and $D(\omega)$ is the matrix associated with the Fourier transform of the noise correlation functions, given by $$D(\omega)={\rm Diag}
[0,(\gamma_m \omega/\omega_m) \coth(\hbar \omega/2k_BT),\kappa,\kappa,\kappa,\kappa].$$
Using the CM $\mathcal{V}^{out}$ one can analyze the entanglement between the three different
bipartitions of the system, when one of the three modes is traced out, and also tripartite entanglement.

\subsubsection{Optomechanical entanglement}

\begin{figure}[htb]
\centerline{\includegraphics[width=0.99\textwidth]{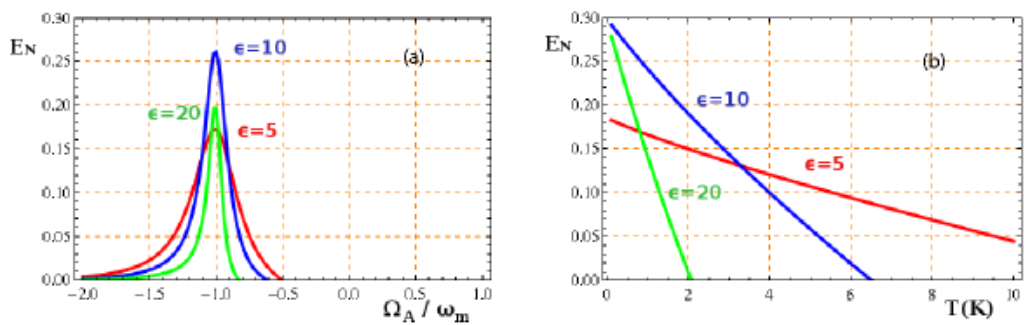}} \caption{Logarithmic negativity $E_{\mathcal{N}}$ of the
bipartite system formed by the mechanical mode and the output of the ``cooling'' mode A. (a) $E_{\mathcal{N}}$ versus the center frequency of the
output mode $\Omega_A/\omega_m$ at three different
values of the inverse detection bandwidth $\epsilon=\omega_m \tau$. As in the single driven mode case (see Fig.~\protect\ref{Stokes_ent}),
entanglement is maximum when the output mode is centered around the
the Stokes sideband $\Omega_A=-\omega_m$. The other parameters are $p_0$, $\kappa=\omega_m$, $\Delta_A=\omega_m$, $\Delta_B=-\omega_m$,
$G_a=0.326 \omega_m$, $G_b=0.302 \omega_m$.
(b) $E_{\mathcal{N}}$ versus the reservoir temperature $T$ when the output mode is centered at the Stokes sideband ($\Omega_A=-\omega_m$)
for the same three different
values of $\varepsilon$.}\label{optomech-cool}
\end{figure}

First of all we consider the entanglement between the output field of the ``cooling mode'' (A) (the one with $\Delta_A >0$)
and the mechanical resonator.
We have seen in Sec.~\ref{singledriven} that this configuration allows to achieve the maximum optomechanical entanglement
in the case of a single driven cavity mode. In fact, when $\Delta \simeq \omega_m$, $G$ is large enough, and the selected output mode
is centered around the Stokes sideband, the entanglement is optimized and it is also robust against temperature (see Fig.~\ref{Stokes_ent}).
Fig.~\ref{optomech-cool} shows that the presence of the second ``heating''
mode B \emph{disturbs} this optimal condition and that $E_\mathcal{N}$ is appreciably lower than the one with only one driven mode. In fact,
we have considered here a similar parameter region, i.e. $p_0$, $\kappa=\omega_m$, $\Delta_A=\omega_m$, $\Delta_B=-\omega_m$,
$G_a=0.326 \omega_m$, $G_b=0.302 \omega_m$. The qualitative behavior of $E_\mathcal{N}$ is
identical to that of the corresponding Fig.~\ref{Stokes_ent}, i.e., $E_\mathcal{N}$ is maximum when the output mode overlaps as best as possible with
the Stokes sideband of the corresponding driving laser, which means centered around $-\omega_m$ and with an inverse bandwidth
$\varepsilon = \omega_m \tau \simeq 10$. However the achievable values of $E_\mathcal{N}$ are significantly \emph{lower}. Fig.~\ref{Stokes_ent}b
shows that, despite the lower values, entanglement is still quite robust against temperature.

\begin{figure}[tb]
\centerline{\includegraphics[width=0.99\textwidth]{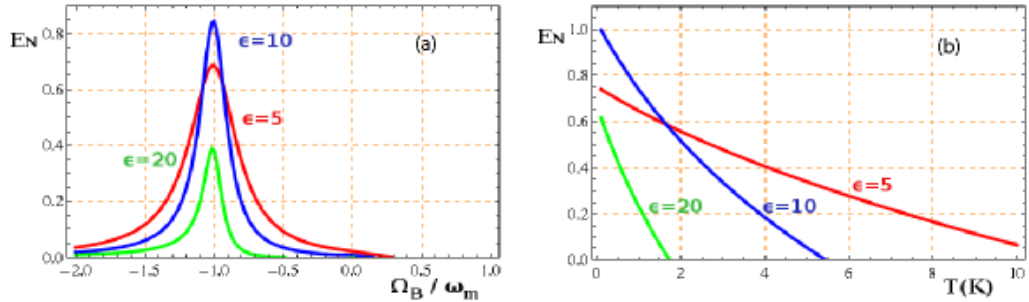}} \caption{Logarithmic negativity $E_{\mathcal{N}}$ of the
bipartite system formed by the mechanical mode and the output of the ``heating'' mode B. (a) $E_{\mathcal{N}}$ versus the center frequency of the latter
$\Omega_A/\omega_m$ at three different
values of the inverse detection bandwidth $\epsilon=\omega_m \tau$. As it happens for the ``cooling'' mode A,
entanglement is maximum when the output mode is centered around the
the Stokes sideband $\Omega_B=-\omega_m$. Parameters are as in Fig.~\protect\ref{optomech-cool}.
(b) $E_{\mathcal{N}}$ versus the reservoir temperature $T$ when the output mode is centered at the Stokes sideband ($\Omega_B=-\omega_m$)
for the same three different
values of $\varepsilon$.}\label{optomech-heat}
\end{figure}

The advantage of the bichromatic driving becomes instead apparent when one considers the bipartite system formed by the resonator
and the output field of the ``heating'' mode (B), the one with $\Delta_B=-\omega_m$. The stationary optomechanical entanglement
can achieve in this case significantly larger values. The results are shown in
Fig.~\ref{optomech-heat} which refers to the same parameters of Fig.~\ref{optomech-cool} and shows the same qualitative behavior: $E_{\mathcal{N}}$
is optimized when the selected output mode well overlaps with the Stokes sideband of the driving laser $\Omega_B=-\omega_m$ and it persists up to
reservoir temperatures of the order of $10$ K. However, $E_{\mathcal{N}}$ is now roughly three times larger than the corresponding value for the
``cooling'' mode. This behavior is different from what is found in Sec.~\ref{singledriven} for a single driven cavity mode, where we have seen
that optomechanical entanglement in the ``heating'' regime of negative detunings is seriously limited by stability conditions. Now, thanks to the
combined action of the two driven modes and to the conditions (\ref{stabtwo}), the system is always stable and the parametric-like process described
in Sec.~\ref{singledriven} is able to generate large and robust entanglement. Therefore we can say that in this bichromatic case,
mode A helps to entangle in a robust way the output of the ``heating'' mode B, by counteracting its instability effect and making the system stable
for any value of the couplings $G_A$ and $G_B$. Notice that in this case, the Stokes sideband of the laser driving mode B is \emph{resonant}
with the cavity, because $\Delta_B=-\omega_m$ implies $\omega_{cB}=\omega_{0B}-\omega_m=\omega_{Stokes}$ and this provides a further reason why
the optomechanical entanglement may become large.

\subsubsection{Purely optical entanglement between output modes}

\begin{figure}[tb]
\centerline{\includegraphics[width=0.99\textwidth]{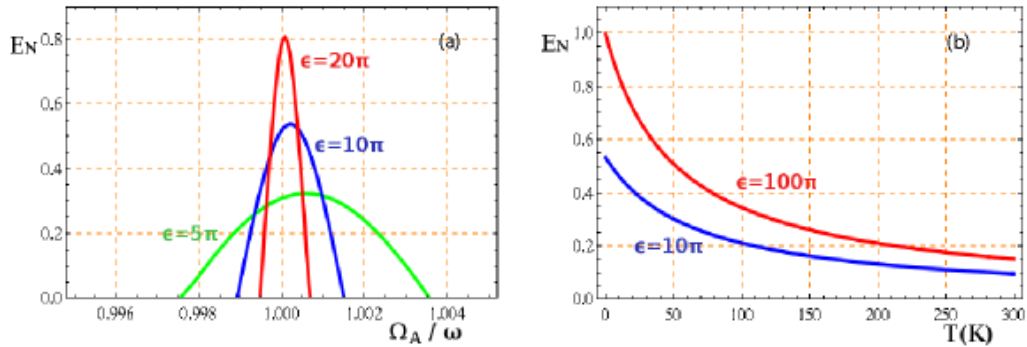}} \caption{Logarithmic negativity $E_{\mathcal{N}}$ of the
bipartite system formed by the output modes associated with the two driven cavity modes. (a) $E_{\mathcal{N}}$
versus the center frequency of the ``cooling'' mode A $\Omega_A/\omega_m$ for a center frequency of the ``heating'' mode fixed at $\Omega_B=-\omega_m$
(Stokes sideband), and at three different values of the inverse detection bandwidth $\epsilon=\omega_m \tau$.
The other parameters are $p_0$, $\kappa=\omega_m$, $\Delta_A=\omega_m$, $\Delta_B=-\omega_m$,
$G_a=1.74 \omega_m$, $G_b=1.70 \omega_m$. (b) $E_{\mathcal{N}}$ versus the reservoir
temperature $T$ when the output of the mode A is centered at the anti-Stokes sideband ($\Omega_A=\omega_m$) and the output of mode B
is centered at the Stokes sideband ($\Omega_B=-\omega_m$),
for two different values of $\varepsilon$.}\label{opto-opto}
\end{figure}

Let us consider now the purely optical entanglement between the two output light beams. As discussed at the beginning of the section,
the possibility to entangle two different output modes of a cavity by means of radiation pressure has been already suggested
in different configurations \cite{wipf,prltelep,giovaEPL01,jopbpir}. We have also seen in Sec.~\ref{singledriven} that this is possible even with a
single driven mode. It is nonetheless interesting to compare the results of Sec.~\ref{singledriven} with the present bichromatic driving case.
The bichromatic case has been already studied in Ref.~\cite{wipf}, which however restricted to the case of output modes with
infinitely narrow bandwidth ($\tau = \infty$) and centered around the driving laser frequency ($\Omega_A=\Omega_B=0$). The general filter function
formalism instead allows us to
consider generic values of $\tau$, $\Omega_A$, and $\Omega_B$. By applying again Eq.~(\ref{Vfinoutbis}) and tracing out now the mechanical mode, we get
the results illustrated in Fig.~\ref{opto-opto}. We have considered a slightly different parameter regime with respect to the previous subsection,
by choosing slightly larger couplings, $G_a=1.74 \omega_m$, $G_b=1.70 \omega_m$, i.e., larger input powers.
Here, the oscillating mirror induces Stokes and anti-Stokes sidebands for \emph{both} driving lasers
and therefore it may be nontrivial to establish which are the most-entangled output modes. Fig.~\ref{opto-opto}(a) shows that the largest
all-optical entanglement is achieved between the \emph{anti-Stokes sideband of the ``cooling'' mode and the Stokes sideband of the ``heating'' beam}.
This is consistent with the results for a single cavity mode, because in both cases the motion of the resonator creates strong
quantum correlations between the scattering of a Stokes and an anti-Stokes photon. Moreover this result can be understood from the fact that
the two sidebands are those which are resonant with the corresponding cavity mode.
Fig.~\ref{opto-opto}(a) also shows that, as it happens in the single cavity mode case, and differently form the optomechanical entanglement,
the all-optical $E_{\mathcal{N}}$ monotonically increases for decreasing detection bandwidths. This is reasonable because
the two output modes are correlated as in two-mode squeezing
which is based on the pairwise correlated production of photons from a pump laser beam via a parametric process. In this case the quantum correlations
are optimally detected when only pairs of photons exactly satisfying the matching
condition $\omega _{s}+\omega _{as}=\omega _{0A}+\omega _{0B}$ are detected, i.e., when $\tau = \infty$.

Fig.~\ref{opto-opto}(b) instead shows the robustness of all-optical entanglement with respect to the reservoir temperature, which is extremely
good: entanglement persists even at room temperature provided that one considers output modes with a sufficiently narrow bandwidth. In this respect,
the bichromatic driving case proves to be more promising than the single driving mode case
(compare Fig.~\ref{opto-opto}(b) with Fig.~\ref{Sidebands_ent}(c)).

Combining all the results of this section, we see that the output modes associated with the two driven cavity modes and the mechanical mode
form a tripartite system in which each bipartite subsystem is entangled. This suggests that a parameter region exists where this
tripartite system is characterized by a fully tripartite-entangled stationary state. This is actually true and it can be checked by
applying the classification criterion of
Ref.~\cite{giedke}, providing a necessary and sufficient criterion for the determination of the entanglement class in the case of tripartite CV
Gaussian states, which is directly computable in terms of the eigenvalues of appropriate test matrices \cite{giedke}.

\section{Cavity-mediated atom-mirror stationary entanglement}

\begin{figure}[tb]
\centerline{\includegraphics[width=0.95\textwidth]{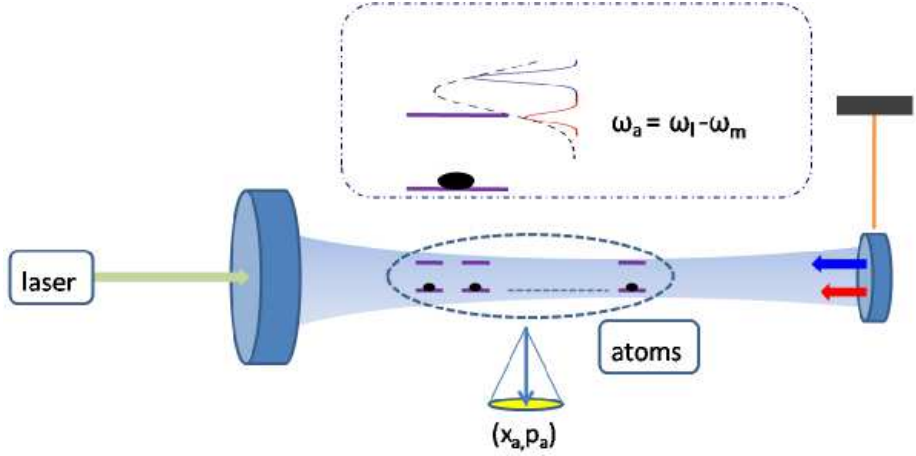}} \caption{Setup for tripartite hybrid
entanglement. An atomic cloud of two-level atoms is placed inside a cavity driven by a laser. As seen in the inset, the atoms are resonant with
the Stokes sideband of the laser. Since this latter sideband is the one carrying most of the optomechanical entanglement, also the atoms and
movable mirror become entangled at the steady state.}\label{AMFscheme}
\end{figure}

A final recent application of optomechanical systems, recently suggested by a number of papers (see Refs.~\cite{fam,ian-ham}), is to couple them also to
atomic ensembles in order to realize new and more flexible CV quantum interfaces. To be more specific, here we consider a hybrid system
comprised of $N_{a}$ two-level atoms of energy splitting $\hbar \omega _{a}$, coupled to an optical cavity, which is in turn coupled to a
mechanical element by radiation pressure. We consider again the steady state of the system and choose a weak-coupling regime where
the atoms and the cavity are far-off resonance (as illustrated by Fig.~\ref{AMFscheme}). The working point for the optomechanical
system is the regime described in the previous section where red-detuned driving of the cavity ensures optimal entanglement between the Stokes
sideband and the mechanical resonator. We show here that when the atoms are resonant with the Stokes sideband of the laser, a regime where both
atoms-mirror bipartite CV entanglement and tripartite CV entanglement can be generated in the steady state, is achieved.

We start from the Hamiltonian of Eq.~(\ref{Ham}) to which we add the
Tavis-Cummings atom-cavity field interaction%
\begin{equation*}
H_{I}=\hbar g\left( S_{+}a+S_{-}a^{\dag }\right) ,
\end{equation*}%
where collective spin operators are defined as $S_{+,-,z}=\sum_{\{i\}}\sigma _{+,-,z}^{(i)}$ for $i=1,N_{a}$ ($\sigma _{+,-,z}$ are the Pauli
matrices) and satisfy the commutation relations $\left[ S_{+},S_{-}\right] =S_{z}$ and $\left[ S_{z},S_{\pm }\right] =\pm 2S_{\pm }$. The
atom-cavity coupling
constant is given by $g=\mu \sqrt{\omega _{c}/2\hbar \epsilon _{0}V}$ where $%
V$ is the cavity mode volume, $\mu $ is the dipole moment of the atomic transition, and $\epsilon _{0}$ is the free space permittivity.

The dynamics of the tripartite system is fairly complicated. However, one can find a regime where a simpler dynamics of three coupled harmonics
oscillators is a good approximation of the system dynamics. To this purpose, we
assume that the atoms are initially prepared in their ground state, so that $%
S_{z}\simeq \left\langle S_{z}\right\rangle \simeq -N_{a}$ and this condition is not appreciably altered by the interaction with the cavity
field. This is satisfied when the excitation probability of a single atom is small. In this limit the dynamics of the atomic polarization can be
described in terms of bosonic operators: in fact if one defines the atomic annihilation operator $c=S_{-}/\sqrt{\left\vert \left\langle
S_{z}\right\rangle \right\vert }$, one can see that it satisfies the usual bosonic commutation relation $[c,c^{\dag }]=1$ \cite{holstein40}. In
the frame rotating at the laser frequency $\omega _{l}$ for the atom-cavity system, the quantum Langevin equations can then be written as
\begin{eqnarray}
\dot{q}& =&\omega _{m}p, \label{QLEs1}\\
\dot{p}& = &-\omega _{m}q-\gamma _{m}p+G_{0}a^{\dag }a+\xi , \\
\overset{\cdot }{a}& =& -(\kappa +i\Delta _{0})a+iG_{0}aq-iG_{a}c+\mathcal{E}_{l}+\sqrt{%
2\kappa }a_{in}, \label{QLEs2}\\
\overset{\cdot }{c}& =&-\left( \gamma _{a}+i\Delta _{a}\right) c-iG_{a}a+%
\sqrt{2\gamma _{a}}F_{c}, \label{QLEs3}
\end{eqnarray}
where $\Delta _{0}=\omega _{c}-\omega _{l}$ and $\Delta _{a}=\omega _{a}-\omega _{l}$ are the cavity and atomic detuning with respect to the
laser, $G_{a}=g\sqrt{N_{a}}$, and $2\gamma _{a}$ is the decay rate of the atomic excited level. The noise affecting the atoms has one
non-vanishing correlation function $\langle F_{c}\left( t\right) F_{c}^{\dagger }\left( t^{\prime }\right) \rangle =\delta \left( t-t^{\prime
}\right) $. We now assume that the cavity is intensely driven, so that at the steady state, the intracavity field has a large amplitude $\alpha
_{s}$, with $|\alpha
_{s}|\gg 1$. However, the single-atom excitation probability is $%
g^{2}|\alpha _{s}|^{2}/(\Delta _{a}^{2}+\gamma _{a}^{2})$ and since this probability has to be much smaller than one for the validity of the
bosonic description of the atomic polarization, this imposes an upper bound to $%
|\alpha _{s}|$. Therefore the two conditions are simultaneously satisfied only if the atoms are weakly coupled to the cavity, $g^{2}/[\Delta
_{a}^{2}+\gamma _{a}^{2}] \ll |\alpha _{s}|^{-2} \ll 1$.

If one is interested only in atoms-mirror entanglement, one could assume a bad cavity limit and adiabatically eliminate the cavity mode
\cite{ian-ham}. However, one can have a more complete information by linearizing the Langevin equations Eqs.~(\ref{QLEs1})-(\ref{QLEs3}) around
the semiclassical steady state and then solving for the exact solution of the $3$-mode system steady state provided by the Lyapunov equation
(\ref{Lyapunov}) \cite{fam}. In fact, owing to the Gaussian nature of the quantum noise terms $\xi $, $a_{in}$ and $F_{c}$, and to the
linearization of the dynamics, the steady state of the quantum fluctuations of the system is a CV tripartite Gaussian state, which is completely
determined by its $6\times 6$ correlation matrix (CM). The corresponding drift matrix after linearization is given by
\begin{equation}
A=%
\begin{pmatrix}
0 & \omega _{m} & 0 & 0 & 0 & 0 \\
-\omega _{m} & -\gamma _{m} & G & 0 & 0 & 0 \\
0 & 0 & -\kappa  & \Delta  & 0 & G_{a} \\
G & 0 & -\Delta  & -\kappa  & -G_{a} & 0 \\
0 & 0 & 0 & G_{a} & -\gamma _{a} & \Delta _{a} \\
0 & 0 & -G_{a} & 0 & -\Delta _{a} & -\gamma _{a}%
\end{pmatrix}%
,
\end{equation}%
while the diffusion matrix is equal to $D=$diag$\left[ 0,\gamma _{m}\left( 2n_{0}+1\right)
,\kappa ,\kappa ,\gamma _{a},\gamma _{a}\right] $. We have solved Eq. (\ref%
{Lyapunov}) for the CM $\mathcal{V}$ in a wide range of the parameters $G$, $%
G_{a}$, $\Delta $ and $\Delta _{a}$. We have studied first of all the stationary entanglement of the three possible bipartite subsystems, by
quantifying it in terms of the logarithmic negativity of bimodal Gaussian states. We will denote the logarithmic negativities for the
mirror-atom,
atom-field and mirror-field bimodal partitions with $E_{ma}$, $E_{af}$ and $%
E_{mf}$, respectively.

\begin{figure}[tb]
\centerline{\includegraphics[width=0.95\textwidth]{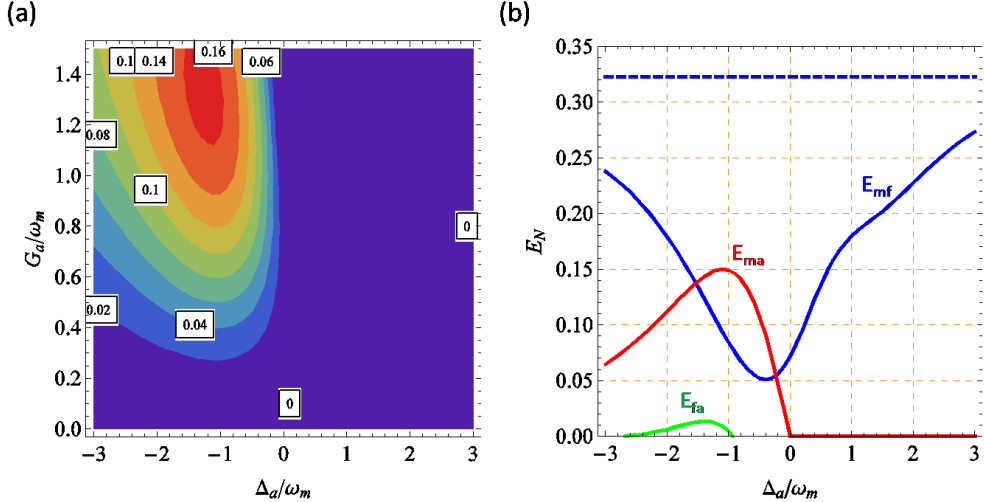}} \caption{Entanglement in the hybrid mirror-atom-field system.
Parameters are $p_{0}$, $\protect\kappa =\protect%
\gamma _{a}=\protect\omega _{m}$, $G=1.3 \protect\omega _{m}$. (a) Contour plot of $E_{\mathcal{N}}$ between mirror and atoms as a function of
$G_{a}/\protect\omega _{m}$ and $\Delta _{a}/\protect\omega _{m}$. The entanglement is optimized for $\Delta _{a}=-\protect\omega _{m}$, i.e.
when the atoms are resonant with the Stokes sideband of the laser. (b) The three bipartite entanglement versus the atomic detuning. The blue
dashed line represents the mirror-field $E_{\mathcal{N}}$ in the absence of atom-field coupling. When the atoms are immersed in the mirror-field
system, the entanglement is redistributed among the three sub-partitions, especially around the regime where $\Delta _{a}=-\protect\omega
_{m}$.}\label{AMF_ent}
\end{figure}

The results on the behavior of the bipartite entanglement $E_{ma}$ are shown in Fig. \ref{AMF_ent}a. Optimization requires, as expected that the
atoms are resonant with the Stokes motional sideband. In Fig.~\ref{AMF_ent}b, the logarithmic negativity of the three bipartitions is plotted
versus the normalized atomic detuning. It is evident that one has a sort of entanglement sharing: due to the presence of the atoms, the initial
cavity-mirror entanglement (represented by the dashed line) is partially redistributed to the atom-mirror and atom-cavity subsystems and this
effect is predominant when the atoms are resonant with the Stokes sideband ($\Delta _{a}=-\omega _{m}$). It is remarkable that, in the chosen
parameter regime, the largest stationary entanglement is the one between atoms and mirror which are only indirectly coupled. Moreover, the
nonzero atom-cavity entanglement appears only thanks to the effect of the mirror dynamics because in the bosonic approximation we are
considering and with a fixed mirror, there would be no direct atom-cavity entanglement. We also notice that atom-mirror entanglement is instead
not present at $\Delta _{a}=\omega _{m}$. This is due to the fact that the cavity-mirror entanglement is mostly carried by the Stokes sideband
and that, when $\Delta _{a}=\omega _{m}$, mirror cavity-cooling is disturbed by the anti-Stokes photons being recycled in the cavity by the
absorbing atoms.

We notice that the chosen parameters correspond to a small cavity mode
volume ($V \simeq 10^{-12}$ m$^3$), implying that for a dipole transition, $%
g $ is not small. Therefore the assumed weak coupling condition $g^{2}/[\Delta _{a}^{2}+\gamma _{a}^{2}] \ll |\alpha _{s}|^{-2} \ll 1$ can be
satisfied only if $g$ represents a much smaller, \emph{time averaged}, coupling constant. This holds for example for an atomic vapor cell much
larger than the cavity mode: if the (hot) atoms move in a cylindrical cell with axis orthogonal to the cavity axis, with diameter $\sim 0.5$ mm
and height $\sim 1$ cm, they will roughly spend only one thousandth of their time within the cavity mode region. This yields an effective $g
\sim 10^4$ Hz, so that the assumptions made here hold, and the chosen value $G_{a}/2\pi =6\times 10^{6}$ Hz can be obtained with $N_a \sim
10^7$. An alternative solution could be choosing a cold atomic ensemble and a dipole-forbidden transition.

The entanglement properties of the steady state of the tripartite system can be verified by experimentally measuring the corresponding CM. This
can be done by combining existing experimental techniques. The cavity field quadratures can be measured directly by homodyning the cavity
output, while the mechanical position and momentum can be measured with the schemes discussed in Sec.~\ref{readout}. Finally, the atomic
polarization quadratures $x$ and $y$ (proportional to $S_x$ and $S_y$) can be measured by adopting the same scheme of Ref.~\cite{sherson}, i.e.,
by making a Stokes parameter measurement of a laser beam, shined transversal to the cavity and to the cell and off-resonantly tuned to another
atomic transition.

\section{Conclusions}

The search for experimental demonstrations of the quantum behavior of macroscopic mechanical resonators is a fastly growing field of
investigation.
Recent experimental results~\cite{naik,arcizet06,gigan06,arcizet06b,bouwm,vahalacool,mavalvala,rugar,wineland,markusepl,sidebcooling,harris,lehnert,vinante,lehnert08,tobias09,markus09}
and theoretical predictions suggest that quantum states of resonators with a mass at the microgram scale will be generated and detected in the
near future. In this chapter we have tried to give an overview of two relevant arguments of this research field: i) cooling to the motional
ground state; ii) the generation of robust entangled steady states involving mechanical and optical degrees of freedom. The latter condition
is the fundamental prerequisite for the eventual integration of micro- and nano-mechanical resonators serving as quantum memories and interfaces
within quantum communication networks.

In the first part of the chapter we have described and compared the two main approaches for cooling micro-mechanical resonators
via radiation pressure coupling to an optical cavity, cold-damping feedback \cite{Mancini98,courty,vitalirapcomm,quiescence02,genes07},
and back-action cooling~\cite{brag,marquardt,wilson-rae,genes07,dantan07,wilson-rae08}. We have adopted a general quantum Langevin treatment which is valid
within the full parameter range of a stable cavity. Both back-action cooling and cold damping feedback are able to cool to the ground state,
even though back-action cooling is preferable for a good cavity ($\kappa < \omega_m$), while cold damping is more convenient for a bad cavity ($\kappa
> \omega_m$).

In the second part of the chapter we have analyzed the entanglement properties of the steady state of the system formed by the
optical cavity coupled to a mechanical element. We have considered two different configurations, with either one or two intensely driven
cavity modes. We have seen that the intracavity mode and the mechanical element can be entangled in a robust way against temperature, and that
back-action cooling is \emph{not} a necessary condition for achieving entanglement. In fact, entanglement is possible also in the
opposite regime of a blue-detuned laser where the cavity mode \emph{drives} and does not cool the resonator.
More generally, the two phenomena are quite independent, and one is
not necessarily accompanied by the other. Cooling is a classical process (even though it can ultimately lead to the quantum ground state), while
entanglement is an intrinsically quantum phenomenon. Moreover, they are optimized in different parameter regimes. In fact, logarithmic
negativity is maximized close to the stability threshold of the system, where instead the resonator is not cooled. We have then focused our study onto the entanglement properties
of the cavity output field, which is the relevant one for quantum
communication applications. We have developed a general theory showing how it is possible
to define and evaluate the entanglement properties of the multipartite system formed by the mechanical resonator and $N$ independent output
modes of the cavity field. We have seen that the
tripartite system formed by the mechanical element and the two output modes centered at the first Stokes and anti-Stokes sideband of the driving
laser (where the cavity output noise spectrum is concentrated) shows robust fully tripartite entanglement. In particular, the Stokes output mode
is strongly entangled with the mechanical mode and shows a sort of entanglement distillation because its logarithmic negativity is significantly
larger than the intracavity one when its bandwidth is appropriately chosen.
In the same parameter regime, the Stokes and anti-Stokes sideband modes are robustly entangled, and the achievable entanglement in the limit of
a very narrow detection bandwidth is comparable to that generated by a parametric oscillators.
These results hold in both cases of single and bichromatic driving of the cavity. In this latter case, entanglement becomes larger
and more robust against temperature under a particular parameter condition in which one mode is driven by a red-detuned laser and the other
one by a blue-detuned laser. In fact, for equal optomechanical couplings and opposite detunings the system is always stable, even for large values
of the intracavity power, and entanglement can persist also at higher temperatures.

Finally we have investigated a possible route for coupling optomechanical devices with atomic ensembles, by showing that if the
atoms are placed inside the optical cavity and tuned into resonance with the Stokes sideband, optomechanical entanglement is optimally
distributed also to the atomic ensemble \cite{fam}. Under these conditions one realizes a strongly coupled system showing robust tripartite entanglement
which can be exploited for the realization of CV quantum interfaces \cite{ian-ham}.

\section{Acknowledgements}

This work has been supported by the European Commission (FP6 Integrated Project QAP, and FET-Open project MINOS), and by INFN (SQUALO project).



\end{document}